# Evidence of *P*-wave Pairing in $K_2Cr_3As_3$ Superconductors from Phase-sensitive Measurement

Zhiyuan Zhang[1, 2, †], Ziwei Dou[1, †], Anqi Wang[1, 2, †], Cuiwei Zhang[1, 2, †], Jiahao Xu[1,2], Yu Hong[1, 2], Xincheng Lei[1, 2], Yue Pan[1, 2], Zhongchen Xu[1, 2], Zhipeng Xu[1, 2], Yupeng Li[1], Guoan Li[1, 2], Xiaofan Shi[1, 2], Xingchen Guo[1, 2], Xiao Deng[1, 2], Zhaozheng Lyu[1], Peiling Li[1],Fanming Qu[1, 2, 3],Guangtong Liu[1, 3], Dong Su[1, 2], Kun Jiang[1, 2, 3], Youguo Shi[1, 2, 3,*], Li Lu[1, 2, 3, *], Jie Shen[1, 3, *], Jiangping Hu[1, 4, 5,*]

[1]Beijing National Laboratory for Condensed Matter Physics and Institute of Physics, Chinese Academy of Sciences, Beijing 100190, China.

[2]University of Chinese Academy of Sciences, Beijing 100049, China.

[3]Songshan Lake Materials Laboratory, Dongguan, Guangdong 523808, China.

[4]Kavli Institute of Theoretical Sciences, University of Chinese Academy of Sciences, Beijing, 100190, China.

[5]New Cornerstone Science Laboratory, Shenzhen 518054, China.

*Corresponding authors. Emails: ygshi@iphy.ac.cn, lilu@iphy.ac.cn, shenjie@iphy.ac.cn, jphu@iphy.ac.cn

†These authors contributed equally to this work.

**Abstract:** *P*-wave superconductors hold immense promise for both fundamental physics and practical applications due to their unusual pairing symmetry and potential topological superconductivity. However, the exploration of the *p*-wave superconductors has proved to be a complex endeavor. Not only are they rare in nature but also the identification of *p*-wave superconductors has been an arduous task in history. For example, phase-sensitive measurement, an experimental technique which can provide conclusive evidence for unconventional pairing, has not been implemented successfully to identify *p*-wave superconductors. Here, we study a recently discovered family of superconductors, $A_2Cr_3As_3$ (A = K, Rb, Cs), which were proposed theoretically to be *p*-wave superconductors. We fabricate superconducting quantum interference devices (SQUIDs) on exfoliated $K_2Cr_3As_3$, and perform the phase-sensitive measurement. It reveals the admixture of 0- and π-phase in these SQUIDs, and we conclude that the existence of the π-phase is in favor of the *p*-wave pairing symmetry in $K_2Cr_3As_3$.

**Main Text:**

## 1. Introduction

The well-established BCS theory adequately explains that in low-$T_c$ superconductor the phonon-mediated many-body electron-electron interaction leads to *s*-wave symmetry order parameter with spin-singlet pairing. However, the nature of the pairing mechanism in high-$T_c$ superconductor remains one of the most intriguing puzzles in condensed matter physics, and promotes the development of methodology for the pairing symmetry experiments. Among these, the phase-sensitive measurements (PSMs), which provide direct evidence for the pairing phase instead of e.g., amplitudes, has settled the key controversy for the unconventional order parameter (*1*). One of the most established PSMs is to construct a Josephson junction (JJ) or SQUID based on the unconventional superconductor with specific geometries and orientations. In this case, the anomaly of the unconventional paring potential, in particular the π-phase across the lobes of the pair wave function with different signs, shifts or deforms the current-phase relation (CPR) of the JJ, and thus is featured by the unusual diffraction pattern of JJ or interference pattern of SQUID. For instance, the minima, instead of the maxima for the conventional superconductivity, of critical current $I_c$ from JJ or SQUID crossing the corner of YBCO appears at zero-field, corresponding to the appearance of *π*-phase shifting between orthogonal direction. The PSM thus contributes mainly to the confirmation of the predominantly $d_{x2-y2}$ wave symmetry in YBCO (*2-4*). Furthermore, other PSMs involving different crystallographic orientations, e.g. the grain boundary, reveal the coexistence of 0- and *π*-phases by the quasi-π-periodic CPR and $I_c$ oscillations of the SQUIDs (*5-8*). This leads to the spontaneous currents and bistable states, providing a strategy to build a so-called “quiet” qubit (*9,10*).

Different from even-parity *s*- or *d*-wave pairing, odd-parity *p*-wave pairing possesses spin-triplet state with spin = 1. It has attracted great attention due to the unique pairing symmetry and parity, as well as the feasibility to host Majorana zero modes in the vortex core and to construct further fault-tolerant topological quantum computation (*11-13*). Though the natural *p*-wave bulk superconductor is rare, there are several candidates, such as non-centrosymmetric superconductor $\beta$-$Bi_2Pd$ (*14*), heavy-fermion metal $UTe_2$ (*15*) and so on. Artificial hybrid systems have also been proposed to realize *p*-wave-like superconductivity, promoting extensive studies about topological materials, semiconductor nanowire with spin-orbit coupling, and ferromagnetic atom coupled to *s*-wave superconductor (*13*). However, no solid-state *p*-wave system has yet been unambiguously identified because the conclusive smoking-gun experimental evidence is still lacking (The superfluid helium-3(*16*) is beyond the system discussed here).

Recently, $A_2Cr_3As_3$ (A = K, Rb, Cs…) family is theoretically predicted to exhibit *p*-wave paring (*17-20*). Due to the reduced dimension in the quasi-one-dimensional (Q1D) structure (*21,22*) and the contribution of 3*d* orbitals of transition metal chromium atoms (*23*), $A_2Cr_3As_3$ family possesses strong electron-electron correlation (*24*) with a superconducting transition temperature $T_c$ ~ 6.1 K at ambience pressure (*21*). A series of experimental explorations have shown its unconventional superconducting signatures (*21,22,25*). The electrical transport measurement reveals a large upper critical field, both parallel and perpendicular to c-axis, exceeding the BCS weak-coupling Pauli limit (*21,26*). NMR measurements suggest $K_2Cr_3As_3$ with spin-triplet superconductivity with the ***d*** vector along c-axis (*25,30*). Besides, the specific-heat data, London penetration depth revealed by a tunnel-diode-based self-inductive technique, and μSR measurements support the existence of unconventional superconductivity for $K_2Cr_3As_3$ (*22,27-29*). However, most of the experimental observations are related to the shape or amplitudes of superconducting gap, which might be modified externally by e.g., defects. Consequently, the phase-sensitive measurements, which are only related to the intrinsic phase of the pairing order parameter, are in demand.

As proposed theoritically (*11*), the ideal phase-sensitive measurement for testing the π-phase in a *p*-wave superconductor is to construct a superconducting loop which connects the opposite facets of the *p*-wave superconductor, as sketched by Fig. S1(A). In this case, there is a π-phase difference between the *s*-*p* and *p*-*s* boundaries, inducing a half magnetic flux quantum in the loop and the minima of oscillating $I_c$ at zero field – the same to the π-phase pattern of "corner SQUID" for *d*-wave-YBCO phase-sensitive measurement (*1*) (also illustrated in Fig. 1G, blue curve). Therefore, we name it as the "polar SQUID". However, in realistic devices even if designed strictly as the theoretical polar SQUID, since the interfaces between *s*-wave (aluminum here) and *p*-wave candidate ($K_2Cr_3As_3$ here) are not perfectly straight (Fig. S1(C)) and atomically-smooth (Fig. S1(D)) proved by images of transmission electron microscope (TEM) (Fig. S2), supercurrent flowing through all the possible facets instead of just along the c-axis provides the superposition of 0- and π-phases (Similar situation was discussed also in *d*-wave YBCO PSM (*8,31*)). Therefore, we term them, as well as the derived configuration in Fig. S1(E), as the "quasi-polar SQUIDs" (see the detailed description for using quasi-polar SOUIDs to test the *p*-wave superconductor here in SM Sec. 4).

We fabricate such quasi-polar SQUID with two aluminum (Al) - $K_2Cr_3As_3$ JJs and measure the interference pattern of critical current $I_c$ as a function of perpendicular magnetic field $B$ at different temperature $T$ below $T_c$ of Al and $K_2Cr_3As_3$. The $I_c(B)$ oscillation displays a traditional 0-phase pattern for the asymmetric SQUID above $T$ = 1 K (*32*), while a cusp-like suppression appears at the maxima of each oscillating period at $T$ < 0.9 K, behaving like an additional modulation superimposed to the normal $B$-dependent periodic $I_c$ oscillation. Both fast Fourier transform (FFT) analysis and quantitative fitting using the model of two-JJ asymmetric SQUID reveal that it is due to the significant contribution from the second-order harmonic component $\sin(2\varphi)$ in CPR, whose amplitude reaches up to above half of the first-order component $\sin(\varphi)$. Such dominant $\sin(2\varphi)$ component beyond the ballistic limit, whose ratio of first ($I_1$)- and second ($I_2$)-order harmonic components ($g = |I_2/I_1|$) is maximally 1/2, for a ordinary JJ is the key observation of this work. It arises from the superposition of 0- and $\pi$-phase supercurrent in the SQUID which reduces the first harmonic while enhances the second harmonic (*5-9*). Therefore, the observation of such dominant $\sin(2\varphi)$ state entails the presence of π-phase supercurrent caused by unconventional pairing. This 0-π phase admixture also leads to the intrinsically bistable states at $\pm\varphi$ and is commonly called $\varphi$-phase state (*33-35*). The accurate value of $\varphi$ depends on $g$ – that is determined by also the weight of 0- and π-phase components. Moreover, we also demonstrate the modulation of the $\varphi$-phase pattern (or $g$) by $B$, and a crossover from 0-dominating to $\pi$-dominating phase state signified by the purely π-periodic $I_c$ oscillation with only $\sin(2\varphi)$ component (*5,8,36*). Furthermore, AC Josephson effect under microwave irradiation manifests the remarkable half-integer Shapiro steps (*36-39*) at zero field and extending to the entire $B$-dependent oscillating period, in agreement with the appearance of the dominant $\sin(2\varphi)$ components (*39*). The above-mentioned measurements also help us to exclude any extrinsic mechanism, such as ferromagnetism, as the source of $\pi$-phase. In conclusion, the PSMs here demonstrate undoubtedly the $\pi$-phase in the pairing wave function of $K_2Cr_3As_3$. Combining with the theoretical predictions (*17-20,23*) and previous experiments (*21,22,25-30*), our results further support this material as a *p*-wave bulk superconductor.

## 2. Measurements and results

<u>Measurement principles of *p*-wave PSM</u>

$K_2Cr_3As_3$ (the crystal structure is sketched in Figs. 1(A) and (B)) possesses double-walled tube along *c*-axis, separated by two kinds of $K^+$ cations with different bonding angles (*17-19*,*22*). Such a typical Q1D structure (*21*,*22*) is helpful for the exfoliation of the materials and the further advanced device fabrication. As shown in Fig. 1(C), the quasi-polar SQUID (also sketched in Fig. S1(E), the reason why we use this configuration is listed in SM Sec. 4) consists of a $K_2Cr_3As_3$ stripe exfoliated from the bulk crystal and Al loop deposited afterwards (the growth and fabrication details are in SM Secs. 1 and 2). The cross-section of $K_2Cr_3As_3$ stripe is more parallelogram- or trapezoid-like due to the lattice structure (*22,40*). It is obvious that the current of the SQUID flows through both $JJ_A$ and $JJ_B$, while $JJ_C$ is connected in series with the SQUID due to the superconducting shortcut by $K_2Cr_3As_3$ below (see more evidence in SM Sec. 12). The zoomed-in sketch in Fig.1(C) highlights the detailed structure of the JJ. The three etched surfaces pointed out by the white arrows are defined as "a/b/c-facets", respectively. The JJ contains the $SiO_2$ barrier at the two opposite "b-facets" reducing the current through these facets (see the details in Figs. S2(E) and S3(A)). In such a quasi-polar SQUID, considering the coupling direction is reversed along c-axis, the phase of supercurrent is sensitive to the π-phase residing in the *p*-wave superconductivity along c-axis. Fig. 1(C) also shows the measurement configuration with standard lock-in technique at a dilution refrigerator. The real device is shown in Fig.1(D). By cooling down from 7 K to ~ 10 mK, we find there are two transitions in the differential resistance d*V*/d*I* vs *T* curve in Fig. 1(E): the first happens at $T$ ~ 4 K and ~ 5 K for Device A and B respectively, indicating $T_c$ of $K_2Cr_3As_3$ stripes, and another at 1.2 K corresponding to $T_c$ of Al for both devices. The superconducting transition slope of $K_2Cr_3As_3$ stripe is not as sharp as that of the bulk crystal, which might be due to the slight aging effect (*21*,*41*). The *R*-*T* curves for all other devices are summarized in Fig. S12 with no clear correlation between the transition slope and the appearance of π-phase supercurrent. Most of our measurement is performed at $T < 1.2$ K to make sure the supercurrent appears in JJs composed by $K_2Cr_3As_3$ and Al.

In JJs with transmission *t*, the CPR acquires higher harmonics and is usually sufficient to consider the first two harmonics (*32*). The conventional supercurrent is in the "0-state" (*4*,*33*), because its phase-dependent supercurrent $I(\varphi)$ is zero and the Josephson energy $U(\varphi)$ ($U(\varphi) \propto \int I(\varphi) d\varphi$) is minimal at $\varphi = 0$ (magenta curves in Fig. 1(F)). Once the unconventional superconductivity is involved (*1*), additional $\pi$ phase appears and results in a "$\pi$-state", which has the minimal Josephson energy at $\varphi = \pi$ (blue curves in Fig. 1(F)) (*4*,*33*). In realistic devices, due to different crystal facets involved in the junctions, both 0- and $\pi$-states could possibly contribute to the total CPR. In such admixture of 0- and $\pi$-states, a "$\varphi$-states" emerges, where its first harmonic can be made arbitrarily small by adjusting the weight of its 0- and $\pi$-phase components (*4*,*33*). In particular, the second harmonic is consequently enhanced and $g = |I_2/I_1|$ may significantly exceed 1/2 which is impossible for individual 0- or $\pi$-state, even reaching the perfect ballistic limit (*5*-*9*). As a result, the CPR for the $\varphi$-state shows characteristic quasi-$\pi$-period oscillation and the Josephson energy $U(\varphi)$ has two degenerate energy minimal points at $\varphi = \pm\arccos(1/2g)$ (black curves in Fig. 1(F)), instead of only one in the 0- or $\pi$-states (magenta and blue curves in Fig. 1(F)). Such two degenerate energy minimal points also result in a spontaneously generated current in the JJ - an important signature for the unconventional superconducting pairing state, and can be harnessed to construct a quiet qubit because of the intrinsic bistable states (*10*). The dominant second harmonic (or $g > 1/2$) is a key signature for $\varphi$-states (*4*,*33*,*35*), accompanied by the appearance of an intermediate maxima (or a kink) between the main maxima in CPR as shown in Fig. 1(F). Such phenomena have been studied extensively in YBCO grain-boundary JJs (*1*,*5*,*6*,*8*) and superconductor-ferromagnet-superconductor (S-F-S) JJs (*36*), as well as considered as a common result of the coexistence of 0- or $\pi$-states (*33*,*36*).

This dominant second harmonic, or the $\varphi$-states, can be directly investigated using a SQUID. As illustrated in Fig. 1(C), the SQUID contains two junctions A and B in parallel, each with the CPR in the generic form of $I_1\sin(\varphi)$ - $I_2\sin(2\varphi)$. The phases across each junction are determined by the magnetic flux threading the loop and the phase drops across the inductance in each arm. Therefore, the total CPR for the SQUID can be modeled as follows (*5,8,36*) ($I_{A1,2}$ /$I_{B1,2}$ are the amplitudes of the first/second harmonics of $JJ_A$/$JJ_B$, respectively):

$$I_{tot}(\varphi_A, \varphi_B) = [I_{A1}\, sin(\varphi_A) - I_{A2}\, sin(2\varphi_A)] + [I_{B1}\, sin(\varphi_B) - I_{B2}\, sin(2\varphi_B)] \text{ (1A)}$$

$$\varphi_A - \varphi_B + 2\pi L[I_A(\varphi_A) - I_B(\varphi_B)]/\Phi_0 = 2\pi\Phi/\Phi_0 \text{ (1B)}$$

For a fixed flux $\Phi$, the critical current $I_c(\Phi) = \max\,[I_{tot}(\varphi_A, \Phi)]$ calculated by equation (1) can be compared directly with the experimental results to extract the harmonics $I_{A1}$, $I_{A2}$, $I_{B1}$, $I_{B2}$. The extraction of the harmonics can be made easier if one junction has much higher critical current than the other, and the shape of $I_c(\Phi)$ is simply the CPR of the weak junction (supposed $JJ_A$ here) (*32*). Therefore, equation (1A) of such an asymmetric SQUID can be simplified as:

$$I_{tot}(\varphi_A, \varphi_B) = [I_{A1}\, sin(\varphi_A) - I_{A2}\, sin(2\varphi_A)] + I_{B1}\, sin(\varphi_B) \text{ (2)}$$

Fig. 1(G) (black curve) illustrates the expected SQUID oscillation when the system is in the $\varphi$-state. Different from the 0- or $\pi$-state SQUID oscillations (magenta and blue curves in Fig. 1(G), respectively), the maximal $I_c$ is not exactly at $\Phi$ = 0 or $\Phi_0/2$ and the location depends on $g$. Moreover, within one period of oscillation $I_c(\Phi)$ two peaks instead of one are clearly visible, which is the key signature of the presence of $\varphi$-state. It becomes perfectly π-periodic when $\sin(\varphi)$ drops to zero and only $\sin(2\varphi)$ exists ($g = \infty$). We note that a large screening factor $\beta = LI_c/\Phi_0 = 0.45$, while shifting the curve in the flux and making the shape of $I_c(\Phi)$ more skewed, does not alter the qualitative features mentioned before (black dashed-curve in Fig.1(G)). We also note that for SQUIDs with high transmission JJs but without φ-state (that is, close but not beyond the ballistic limit, $g \leq 1/2$) cannot produce such quasi-π-periodic oscillations since it includes all even and odd higher harmonics (*42*), while the φ-state enhances the even harmonics and suppresses the odd harmonics.

As shown in SM Secs. 4 and 5, due to the involvement of different facets, both 0- and π-state supercurrents exist in the *p*-wave section between two junctions of the quasi-polar SQUID here. We demonstrate in SM Sec. 3, the superposition of both 0- and π-state is exactly equal to a SQUID owning a JJ with the admixture of 0-state and π-state illustrated by equation (1). Such a superposition results in the φ-state SQUID which suppresses the first harmonic while enhances the second harmonic. Therefore, the π-state (necessarily included in φ-state) can be experimentally detected by the enhanced second harmonics and the characteristic quasi-π-periodic oscillations in $I_c(\Phi)$ of the quasi-polar SQUIDs.

<u>Temperature dependence of π-phase supercurrent</u>

To directly test whether such $\varphi$-state (or more precisely π-state) exists in Device A, we measured the $I_c$ vs $B$ oscillation of the SQUID at different $T$. The interference pattern at high $T$ ($T$ = 1 K in Figs. 2(A), (B)) exhibits a slightly tilted and shifted oscillation with respect of $B$ and current bias $I_{DC}$ due to the inductance of the SQUID (see SM Sec. 9). The measured period of the $I_c$ oscillation is around 0.058 mT, which translates to an effective SQUID area of 34.5 μm$^2$, close to the designed geometry around 32 μm$^2$ (SM Sec. 2). Gradually, a dip appears at the maxima and the oscillation splits into quasi-π-periodic oscillation when $T$ < 0.9 K. The quasi-π-periodic oscillation at 0.1 K is directly confirmed in the linecut series shown in Fig. 2(C). To get the quantitative result, we calculate the Fourier expansion of the $I_c$ oscillation and plot the coefficients $F_{1,2}$ vs $T$ in Fig. 2(D) (see Fig. S7 for details of FFT and Fig. S8 for the full Cooldown 2 data with higher temperature resolution). These results give a general description of the amplitudes of the first- and second-harmonic terms, respectively (*6*). Apparently, $F_2$ increases drastically with decreasing $T$ and reaches above half of $F_1$ around $T$ = 0.8 K (see $F_1/F_2$ in Fig. 2(F)), which is consistent with the appearance of quasi-π-periodic feature in the $I_c(B)$ curves in Fig. S8.

Moreover, we fit the experimental data with the two-JJ asymmetric SQUID model as discussed in equation (2) (see the typical fitting results as shown by the yellow curves in Figs. 2(A) and Fig. S8). The resulted $T$-dependent coefficients for the two JJs respectively are listed in Fig. 2(E), with an additional value of $I_c$ for $JJ_B$ (the strong junction) compared to the Fourier coefficients. (The detailed fitting results for each $T$ and discussion are in SM Sec. 9, with $L$ = 88 pH and another two parameters $\Delta\varphi_2$ and $\Delta\varphi_S$ probably arising from the spontaneous current, which have been discussed in YBCO and S-F-S $\varphi$-junction (*11,33,34,43-47*).) Except that, the first and second-harmonic terms $I_{A1,\ A2}$ have the similar value and $T$-dependence to $F_{1,2}$, showing that these two quantitative analysis methods are reasonable and accurate. Based on these, there are two conclusions here: 1. $g$ > 1/2 when $T$ < 0.9 K (Fig. 2(F)), revealing the coexistence of 0 and $\pi$-phase components in the SQUID at low $T$; 2. For $JJ_A$ (the weak junction), $I_{A1}$ decreases and $I_{A2}$ increases, which means the $\pi$-phase component becomes larger, with decreasing $T$. Such strong $T$-dependent behavior is commonly observed in YBCO-based JJ and is possibly related to the local properties of the devices (*8*), such as the distribution of current along different facets (*47*), the transparency-dependent coupling to different wavefunction (*1,31*), and the varying stress of two materials in the junctions (*48*). Meanwhile, we emphasize that we find similar behavior for different cooldown processes with close-to-zero magnetic field to exclude the effect of random vortex pinning (see Cooldown 1 and 2 in Fig. 2 and Fig. S8). We also reproduce such $\varphi$-state SQUIDs in seven devices (SM Sec. 10 and Figs. S10 and S11). All these prove the robustness of 0- and $\pi$-phases in the quasi-polar SQUID, which is consistent with the fact that the predicted $p$-wave pairing results in 0- or $\pi$-phase along the different directions, for example, parallel to $ab$-plane (not reversed by 180°) or $c$-axis as pointed out by the magenta/blue arrows in Fig. 1(C) (*17-20,23*). Of course, a mixture of $s$- and $p$- wave pairing in $K_2Cr_3As_3$ would also result in such a $\varphi$-state SQUID.

<u>Magnetic-field dependence of π-phase supercurrent</u>

More interestingly in the $\varphi$-state SQUID, the weight of 0- and $\pi$-phase components is able to vary the sign of the first-harmonic term through a sweet spot with equal 0- and $\pi$-phase components, where there is no first term and a purely π-periodic, instead of 2$\pi$-periodic,CPR. By applying stronger $B$ to Device A, we find the oscillation pattern changes drastically as other YBCO $\varphi$-junctions show (*5,8*) and one-peak reappears at high magnetic field as illustrated in Fig. 3(A) (also see the zoomed-in oscillation at different $B$ in Fig. 3(B)). The aforementioned π-periodic structure shows up around $B$ ~ 0.5 mT (the arrow in Fig. 3(A), upper plot). Also, a $I_c$ dip in the Fraunhofer pattern appearing at non-zero field has also been observed, similar to that in $d$-wave YBCO $\varphi$-junctions (*8,34,49*). We also notice the pattern is always point symmetric with respect of $B$ = 0 obeying the Onsager relation, regardless of the detailed 0- and π-phase mixture (see more data at different $T$ in Fig. S6) (*8,34,49*). Similarly, we use both the FFT and SQUID model analysis for each period at different $B$ (Figs. 3(B),(C) and Fig. S9), and show the detailed results in Figs. 3(D)- (F). Indeed, the two kinds of results match each other, and $F_1$, $I_{A1}$, $F_2/F_1$ and $I_{A2}/I_{A1}$ all have a kink at $B$ ~ 0.5 mT, consistent with the sign change of the first term at the sweet spot (The gradual evolution of the first and second terms from FFT is clearly illustrated in Fig. 3(C)). That is, below/above $B$ = 0.5 mT, the 0-phase component is bigger/smaller than $\pi$-phase component, and the first term has positive/negative sign. All these behaviors show that the sweet spot represents a dominant 0- to $\pi$-phase transition. In principle, such transition may originate from the supercurrent flowing in both c-axis and ab-plane. However, the illustration in Fig. 3(A) shows that due to the Meissner effect, the supercurrent flowing in c-axis is predominantly exposed and thus modulated by the magnetic field (see Fig. S3(B) for more discussions). This thus might indicate that the $\pi$-phase supercurrent mainly comes from c-axis. Furthermore, we perform more measurements about $B$-dependent pattern with different sweeping directions (listed in SM Sec. 11), showing obviously no hysteresis and thus ruling out ferromagnetism as the source of the observed $\pi$-phase in Al-$K_2Cr_3As_3$ SQUIDs.

Shapiro step measurement

Such a quasi-π-periodic $I_c$ oscillation could also arise from the beating of the two 2π-periodic oscillations with different periods (*50*), apart from the enhanced second harmonics ($g$ > 1/2) due to the admixture of 0- to $\pi$-phases. To further confirm whether the enhanced second harmonics indeed exists, we measure the *I-V* curves under microwave irradiation, where half-integer Shapiro steps should appear in this case. Here, the phase-locked response to the irradiation at a given frequency produces a series of quantized steps in voltage (or Shapiro steps) as a result of the AC Josephson effect. Therefore, for the JJ with the CPR such as $I_1\sin(\varphi) - I_2\sin(2\varphi)$, the first harmonic produces the integer Shapiro steps $V_n = nhf/2e$ (n = ±1, ±2, ±3,...), while the second harmonic produces the half-integer Shapiro steps $V_n = nhf/4e$ (n = ±1, ±2, ±3,...) (*39*), which are what we intend to observe. In the past, such method has been applied to investigate the half-integer harmonic contents in systems such as $\varphi$-junction state in *d*-wave JJs (*5,8,37*), ferromagnetic JJs (*36,38*) and in the newly discovered twisted BSCCO JJs (*43*), high transmission SNS junction using high-mobility graphene or semiconductor (here $g \leq 1/2$) (*42,51,52*) and enhanced higher harmonics due to non-equilibrium process (*53*). We note that in our overdamped junctions (see Figs. S18 and S19) the association between half-integer Shapiro steps and second harmonic of CPR is straightforward without complications due to the chaotic phase-lock dynamics (*39*).

In the experiment, it is more common to measure the Shapiro map, tracing the dependence of the differential resistance d$V$/d$I$ as a function of the a.c. and d.c. current drive ($I_{AC}$ and $I_{DC}$ respectively). Fig. 4(A) shows a typical theoretical Shapiro map calculated by the widely used RSJ model using $g$=1.2 same as the values extracted in Fig. 2(F) (detailed in SM Sec. 13), and the 1/2 step emerges between the two sequential integer steps, corresponding to the dominant second harmonic in $\varphi$-state. In Fig. 4(B), we measure such Shapiro map in Device A at $B$ = 0, $T$ = 0.1 K and $f$ = 4.69 GHz (or $\nu = f/f_J \approx 0.35$, $I_c$ = 4.6 μA, $R_N$ = 6.0 Ω, $f_J = 2eI_cR_N/h$ being the Josephson frequency (*39*)). The evolution of the 1/2 step with power is measured only up to 20 dBm due to the maximal power provided by the microwave source. The Shapiro map shows the qualitatively similar features as the theoretical map in Fig. 4(A), in particular a clear 1/2 step (red arrow). The high $I_{DC}$ region of the map is somewhat distorted by an addition JJ ($JJ_C$) outside the SQUID which we carefully distinguish in Fig. S17, but does not affect the qualitative feature at low $I_{DC}$ region. By integrating the measured d$V$/d$I$($I_{DC}$) and obtaining the histogram of the *I-V* curves for each $P$ in Fig. 4(C), the steps are better illustrated and the clear peak at 1/2 step ($V$ = (1/2)nh$f$/2e, magenta arrow) is seen. As shown in Fig. 4(D) taken at different $T$, such 1/2 step at zero magnetic field persists at low temperatures and becomes blurred above 0.7 K, which is similar to the temperature range where $g$ decreases from above to below 1/2 with higher $T$ in Fig. 2(F) and thus proves they share the same origin.

By fixing the microwave power and sweeping the magnetic field, we can further observe the evolution of the half-integer step as a function of the flux. Fig. 4(E) is the $dV/dI(I_{DC}, B)$ data taken without microwave irradiation at $T$ = 0.1 K, showing clearly the quasi-π-periodic feature. By introducing the microwave with $f$ = 4.69 GHz ($P$ = 16 dBm, corresponding to the magenta vertical dashed line in Fig. 4(B)), the 1/2 step is continuous throughout the whole period of the SQUID (Figs. 4(F),(G)). In addition, another similar device (Device B) shows the quasi-π-periodic oscillation without microwave only for negative bias, while the 2π-periodic oscillation shows for positive bias (Fig. 4(H)). Using the microwave with similar $\nu = \nu = f/f_J \approx 0.35$ we reproduce the clear 1/2 step. Importantly, in Figs. 4(I),(J), such 1/2 step is continuous throughout the whole period only for negative bias, while it only appears around the flux $\Phi \approx \Phi_0/2$ for positive bias (more data see Fig. S20). Therefore, there is a close correlation between the quasi-π-periodic oscillation and the continuous 1/2 step throughout the whole period.

**3. Discussion for alternative mechanisms**

We note several scenarios where the 1/2 step may appear without $\varphi$-state ($g \leq 1/2$), which we exclude as follows. First, a high frequency comparable to $f_J$ or a high power above the first node will induce half-integer Shapiro step even when $g \leq 1/2$, as both $f$ and $P$ will enhance the width of the 1/2 step (*39*). However, in Fig. 4(A) and Fig. S19, we show that for the low microwave frequency $\nu = f/f_J = 0.35$ used here the half-integer steps at low power before the first node occurs only for $g > 1/2$. Second, unusual 1/2 steps could be caused by the microwave-induced non-equilibrium and still survive at higher $T$. This is inconsistent with the 1/2 steps here which disappear at higher temperature accompanied by the degradation of quasi-π-periodic oscillation (*53*). Finally, the half-integer Shapiro step at low frequency and power may also arise in SQUIDs without $\varphi$-state. However, it occurs only around the flux $\Phi \approx \Phi_0/2$ where the first harmonic of the two JJs is much reduced due to destructive interference, and thus $g$ becomes larger than 1/2 (*54*). On the contrary, the 1/2 step in Fig. 4 is continuous throughout the whole period in two devices, in particular at zero flux, thus excluding such interference mechanism. This is clearly demonstrated by the comparison between positive and negative bias in Device B, which shows that quasi-π-periodic feature and the half-integer Shapiro steps in the whole period appear/disappear simultaneously for the negative/positive bias (Figs. 4(I),(J)), demonstrating the close correlation between them. This further eliminates the interpretation of the quasi-π-periodic $I_c(B)$ as the beating between two 2π-periodic oscillating channels (*50*), since the beating effect does not induce large second harmonics. Besides, quasi-π-periodic oscillations also appear in double-SQUIDs due to the above-mentioned destructive interference at finite magnetic field (55), which is totally different from the quasi-π-periodic oscillation at zero-field here. The quasi-π-periodic resistance oscillation previously observed in ordinary 0-SQUID (56) should also not be confused with the quasi-π-periodic oscillation of $I_c$ here. Therefore, the measured quasi-π-periodic $I_c(B)$, as well as the robust half-integer Shapiro steps, can only be caused by the co-existence of 0- and π-phases supercurrent.

Meanwhile, the unconventional π-phase component may originate from several mechanisms beside $p$-wave superconductivity, such as ferromagnetism (*36*) and other exotic pairing states (*22,29,57*). We can exclude the former as follows: First, previously experiments observed the non-static spin-fluctuations with the short-range spin configurations flipping on the atomic scale. However, this does not produce a static long-range ferromagnetic order and thus the π-phase supercurrent in our micron-sized device either. Second, as more data shown in SM Sec. 11, no hysteresis or mirror-symmetry in $I_c(B)$ is observed between up and down sweeps of $B$, thus demonstrating the absence of any ferromagnetism in the system. Besides ferromagnetic states, exotic s-wave states, such as a variety of sign changed s-wave states proposed for iron-based superconductors (*58*), cannot be applicable to the π-phase supercurrent observed in our device because the sign change is in the momentum space and does not exist in real space. Furthermore, exotic inter-band pairings in $K_2Cr_3As_3$ are extremely difficult to form since the dispersions and the Fermi surfaces of these three bands in $K_2Cr_3As_3$ are very different (*17,19*). Therefore, we determine that the unconventional *p*-wave superconductivity is the cause for the π-phase component.

As for the type of *p*-wave pairing, we have discussed in SM Sec. 5. In these quasi-polar devices, the phase of the supercurrent is less sensitive to *p*-wave pairing along the a/b-axis than the c-axis and thus the π-phase supercurrent might be mainly from c-axis, which is supported in particular by the strongly *B*-modulated 0 and π-components in Fig. 3. More data show the comparison between a quasi-polar SQUID along the c-axis, a quasi-polar SQUID along the a/b axis, and an edge SQUID fabricated on the same $K_2Cr_3As_3$ crystal (Fig. S4). Around zero field, the quasi-π-periodic $I_c(B)$ only appears in the quasi-polar SQUID along the c-axis, which also indicate the π-phase from c-axis is more predominant than that from the a/b-axis. However, we still need more evidence to prove the type of *p*-wave pairing in $K_2Cr_3As_3$, which is beyond the scope of this manuscript. We reproduce the quasi-π-periodic $I_c(B)$ in seven quasi-polar SQUID devices (Devices A-G, Figs. S10 and S11), while five more devices do not show such feature (H1-H5, SM Sec. 10). While the devices are fabricated by similar procedures, the degradation with the dissipation of K element (shown by EDS in Fig. S13), the different geometry of $K_2Cr_3As_3$ stripes, the anisotropic etching, and the unintentional deposition of $SiO_2$ sputtered from the substrate by the etching ions (*59*) (Fig. S2) result in a different quality of the Al-$K_2Cr_3As_3$ interfaces, which affects *s*-*p* coupling (*60*) and the weight of the π-phase component.

## 4. Conclusion

In conclusion, we have measured quasi-π-periodic $I_c(B)$ oscillation at low temperature and attribute this to the enhanced second-order harmonic component $\sin(2\varphi)$. This anomalous $\sin(2\varphi)$ has also been proved by the significant half-integer Shapiro steps appearing in the entire *B*-dependent period and reveals the coexistence of 0- and $\pi$-phases. Apart from temperature, magnetic field also modules the weight of 0- and $\pi$-phase components, and induces a dominant 0- to $\pi$-phase transition. By varying the sweeping direction of the magnetic field, we can exclude any ferromagnetic component as the source of the $\pi$-phase here. The appearance of $\pi$-phase, which is reproducible in multiple devices, demonstrates the sign change in the superconducting pairing order parameter, and is consistent with predicted *p*-wave pairing in $K_2Cr_3As_3$. Such a $\varphi$-phase SQUID pattern is possibly more convincing than that of purely $\pi$-phase SQUID. The reason is that it superimposes a pattern on the original 0-phase SQUID oscillation, instead of shifting the entire pattern by half of *B* period, which might arise from the residual magnetic field in the magnet, the vortex pinning in the device or the inductance of the loop. This Q1D *p*-wave superconductor with the advantages of desirable superconducting transition temperature at ambient pressure and compatibility of the state-of-the-art lithography technique provides an ideal platform not only to perform the phase-sensitive measurement for a *p*-wave superconductor, but also to develop novel superconducting devices and qubits, e.g. the long-sought quiet qubit and topological qubit.

**Acknowledgments:**

**Funding:**

The work of J.S., L.L., F.Q. and G.L. was supported by:

Beijing Natural Science Foundation (Grant No. JQ23022)

Strategic Priority Research Program B of Chinese Academy of Sciences (Grant No. XDB33000000)

Beijing Nova Program (Grant No. Z211100002121144)

National Natural Science Foundation of China (Grant Nos. 92065203 and 12174430)

Synergetic Extreme Condition User Facility (SECUF)

The work of K. J. and J. H. was supported by:

Ministry of Science and Technology (Grant No. 2022YFA1403900)

National Natural Science Foundation of China (Grant No. NSFC-11888101, No. NSFC-12174428, No. NSFC-11920101005)

Strategic Priority Research Program of the Chinese Academy of Sciences (Grant No. XDB28000000, XDB33000000)

New Cornerstone Investigator Program

Chinese Academy of Sciences Project for Young Scientists in Basic Research (2022YSBR-048)

The work of Y.S. was supported by:

National Natural Science Foundation of China (Grants No. U2032204)

Informatization Plan of Chinese Academy of Sciences (CAS-WX2021SF-0102)

Strategic Priority Research Program of the Chinese Academy of Sciences (Grants No. XDB33030000).

**Author contributions:**

J.P.H. and J. S. conceived and designed the experiment.

Z.Y.Z., Z.W.D., A.Q.W., J.H.X., Y.H., Y.P.L., G.A.L., X.F.S., X.C.G., and X.D. fabricated devices and performed the transport measurements, and discussed with Z.Z.L., P.L.L., F.M.Q., G.T.L, supervised by L.L., J.S. and J.P.H..

C.W.Z., Z.C.X., and Y.G.S. grew $K_2Cr_3As_3$ single crystals.

X.C.L., Y.P., and D.S. performed the STEM measurements.

Z.Y.Z., Z.W.D., and Z.P.X. performed the theoretical analysis, supervised by K.J., J.S., and J.P.H..

Z.Y.Z., Z.W.D., A.Q.W., J.S., and J.P.H. analyzed the data and wrote the manuscript, with input from all authors.

**Competing interests:** Authors declare that they have no competing interests.

**Data and materials availability:** All data needed to evaluate the conclusions in the paper are present in the main text and/or the supplementary information. Raw data generated in this study are available from the corresponding author upon reasonable request.

**Supplementary Materials**

Supplementary Text

Figs. S1 to S20

Tables S1

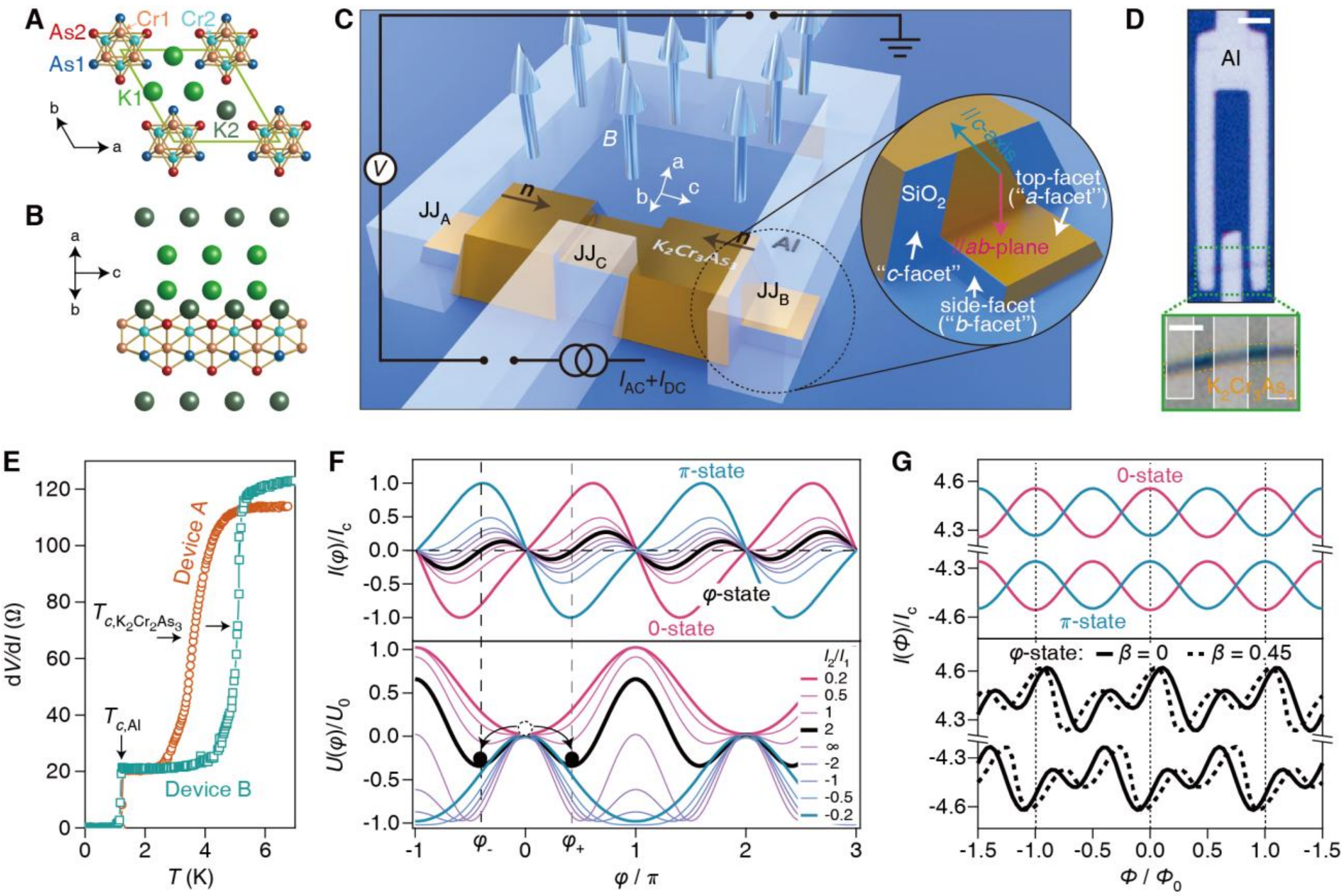

**Fig. 1. Phase-sensitive measurement of $K_2Cr_3As_3$ SQUID.** Crystal structure of $K_2Cr_3As_3$: **(A)** top view. **(B)** side view. **(C)** $K_2Cr_3As_3$ SQUID with Al contacts and measurement configuration. Inset: zoomed-in JJ with a thin layer of sputtered dust of $SiO_2$ onto the etched surfaces. The supercurrent flowing via the ab-plane and c-axis are marked in magenta and blue, respectively. **(D)** Optical image of the real device (scale bar: 2 μm) with the zoomed-in junction area (green rectangle, scale bar: 1 μm). **(E)** $R = dV/dI$ versus $T$ showing a slow superconducting transition around 4 K and 5 K for $K_2Cr_3As_3$ in Device A and B and a sharp transition at 1.2 K for Al in both Devices. **(F)** CPR (upper) and Josephson energy (lower) for various $I_2/I_1$. When $g = |I_2/I_1| < 0.5$, the system only has one stable state at either $\varphi = 0$ (0-state) or $\varphi = \pi$ (π-state). When $g > 0.5$, the system has two stable states at $\varphi_{\pm} = \pm\arccos(1/2g)$ ($\varphi$-state) . **(G)** $I_c$ versus $\Phi$ of the SQUID. Magenta: 0-state, blue: π-state, black: $\varphi$-state. The π-periodic structure only appears in $\varphi$-state SQUID. Dashed line: inductance effect increasing the misalignment between $I_{c+,-}$. $\beta = LI_c/\Phi_0 = 0.45$.

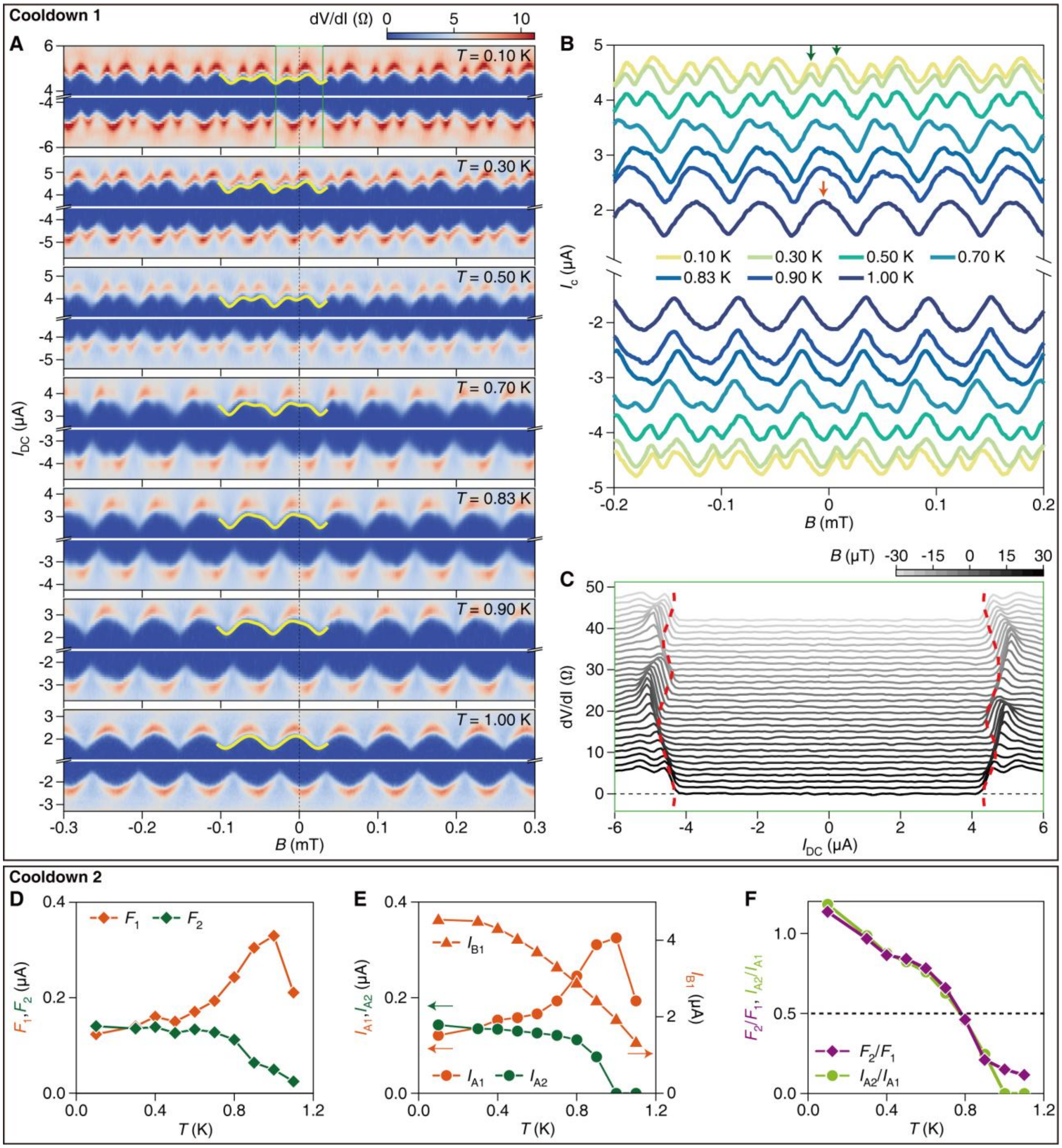

**Fig. 2. Flux-dependent critical current at different temperatures.** Cooldown 1: **(A)** d$V$/d$I$($I_{\mathrm{DC}}$, $B$) with $T$ = 0.1, 0.3, 0.5, 0.7, 0.83, 0.9, 1.0 K. For each temperature, $I_{\mathrm{DC}}$ sweeps in the positive/negative direction. The quasi- "π-periodic" $I_{\mathrm{c}}$ structure in each period is obvious at for $T \leq 0.83$ K (dominated by the second harmonic) (see text). At $T$ = 1.0 K, this is reduced to the single peak shape (dominated by the first harmonic). The fitting results of $I_{\mathrm{c}}(B)$ are overlapped for all the figures. **(B)** Extracted $I_{\mathrm{c}+,-}$ versus $B$ for all the temperatures. **(C)** d$V$/d$I$ versus $I_{\mathrm{DC}}$ within a period between B = -0.03 mT to 0.03 mT (taken from (A) within the green rectangle) with the same offset between sequential curves for clarity. The extracted $I_{\mathrm{c}}$(B) same as (B) is overlaid (red dashed lines). $T$ = 0.1 K. Cooldown 2: **(D)** FFT components ($F_1$, $F_2$) of $I_{\mathrm{c}}(B)$ versus $T$ with better temperature

resolution (see text). **(E)** $I_{A1}$, $I_{A2}$ for $JJ_A$ and $I_{B1}$ for $JJ_B$ versus $T$ from the fitting (see text). **(F)** $F_1/F_2$ (purple) and $g = I_{A2}/I_{A1}$ (prasinous) versus $T$. $g$ exceeds 1/2 below 0.8 K.

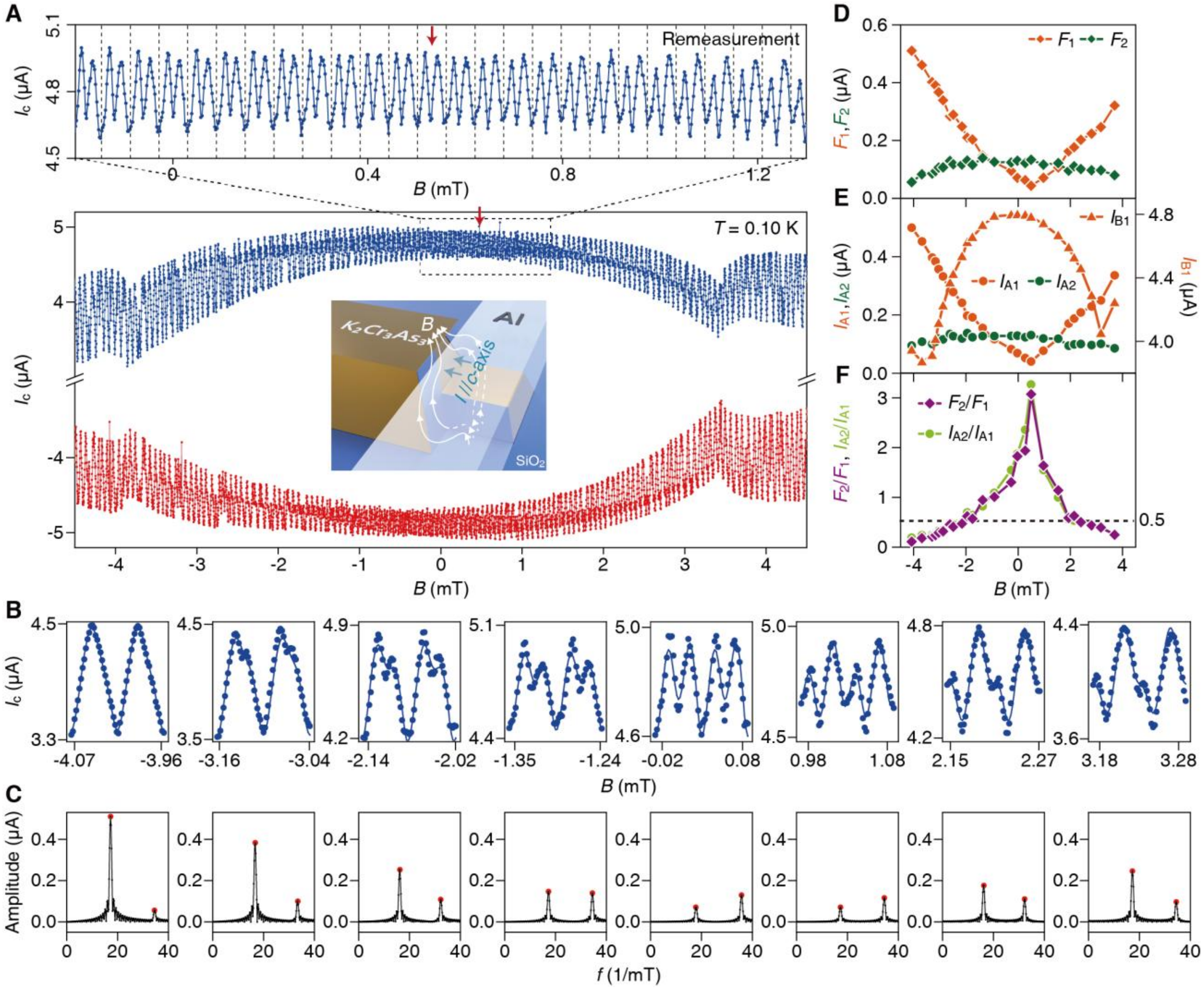

**Fig. 3 Flux-dependent critical current at wide range of magnetic field. (A)** Positive and negative critical current $I_{c+,-}(B)$ at $T = 0.1$ K, showing π-to-2π periodic transition. Upper plot: the remeasured zoomed-in plot across the transition region for $B$ from -0.2 mT to 1.3 mT. Red arrows: $B$ with the largest ratio between the second and first harmonics. More data is in Fig. S6. Inset: The Illustration of distorted magnetic field lines wrapping around $K_2Cr_3As_3$ at the interface between the Al film and $K_2Cr_3As_3$. **(B)** measured $I_c(\Phi)$ at different $B$ (blue dots) with the fitting curves of the SQUID model (blue lines). **(C)** Corresponding FFT amplitude spectrum of $I_c(\Phi)$, showing the first two harmonics. The first and the second FFT components $F_1$ and $F_2$ are extracted at the red points. **(D)** The first and the second FFT components ($F_1$, $F_2$) versus $B$. **(E)** $I_{A1,2}$ and $I_{B1}$ from the fitting versus $B$. **(F)** $F_1/F_2$ (purple) and $g = I_{A2}/I_{A1}$ (prasinous) versus $T$. $g$ exceeds 1/2 for $|B|$ within 2 mT. See SM. Sec. 6 for the fast measurement technique used to measure the wide range $I_c(B)$.

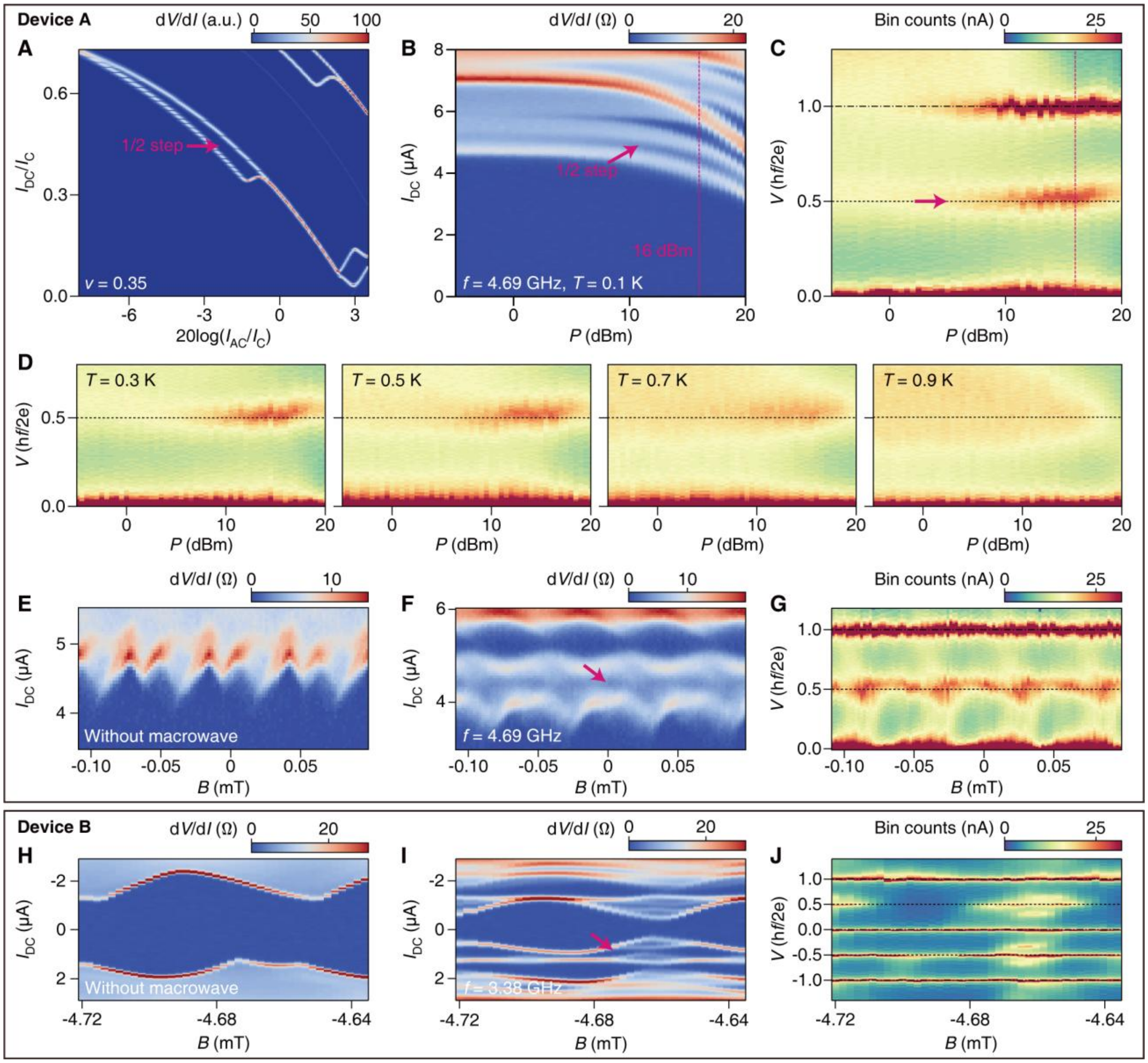

**Fig. 4. Dominant second harmonic demonstrated with fractional Shapiro step.** Device A: **(A)** Calculated Shapiro maps showing half-integer steps (magenta arrow) between subsequent integer steps for $g$ = 1.2 and $v$ = 0.35 (see text). **(B)** $R$ = d$V$/d$I$ versus $I_{DC}$ and $P$. $f$ = 4.69 GHz ($v \approx 0.35$), $B$ = 0 T, $T$ = 0.1 K. The 1/2 step is seen, agreeing with (A). **(C)** Histogram from $I$-$V$ curve versus $V$ and $P$ showing clear 1/2 step at zero field. **(D)** Histograms at $B$ = 0 T from $T$ = 0.3 to 0.9 K.The 1/2 step clear below 0.7 K diminishes at higher temperatures. **(E)** d$V$/d$I$($I_{DC}$, $B$) at $T$ = 0.1 K without irradiation. **(F,G)** d$V$/d$I$($I_{DC}$, $B$) and histogram. $P$ = 16 dBm (vertical dashed line in (B)). The 1/2 step (the magenta arrow) is continuous throughout the whole period. Device B: **(H)** d$V$/d$I$($I_{DC}$, $B$) without microwave showing quasi-π-periodic (2π-periodic) oscillations for negative (positive) bias around -4.68 mT. $T$ = 96 mK. **(I, J)** d$V$/d$I$($I_{DC}$, $B$) and histograms with $f$ = 3.38 GHz ($v \approx 0.35$) and $P$ = -10 dBm. The 1/2 step (magenta arrow) continuous over the whole period only for negative bias with quasi-π-periodic oscillations.

# Supplementary Materials for

## Evidence of *P*-wave Pairing in $K_2Cr_3As_3$ Superconductors from Phase-sensitive Measurement

Zhiyuan Zhang[1, 2, †], Ziwei Dou[1, †], Anqi Wang[1, 2, †], Cuiwei Zhang[1, 2, †], Jiahao Xu[1,2], Yu Hong[1, 2], Xincheng Lei[1, 2], Yue Pan[1, 2], Zhongchen Xu[1, 2], Zhipeng Xu[1, 2], Yupeng Li[1], Guoan Li[1, 2], Xiaofan Shi[1, 2], Xingchen Guo[1, 2], Xiao Deng[1, 2], Zhaozheng Lyu[1], Peiling Li[1],Fanming Qu[1, 2, 3],Guangtong Liu[1, 3], Dong Su[1, 2], Kun Jiang[1, 2, 3], Youguo Shi[1, 2, 3,*], Li Lu[1, 2, 3, *], Jie Shen[1, 3, *], Jiangping Hu[1, 4, 5,*]

**Affiliations:**

[1]Beijing National Laboratory for Condensed Matter Physics and Institute of Physics, Chinese Academy of Sciences, Beijing 100190, China.

[2]University of Chinese Academy of Sciences, Beijing 100049, China.

[3]Songshan Lake Materials Laboratory, Dongguan, Guangdong 523808, China.

[4]Kavli Institute of Theoretical Sciences, University of Chinese Academy of Sciences, Beijing, 100190, China.

[5]New Cornerstone Science Laboratory, Shenzhen 518054, China.

Corresponding authors: ygshi@iphy.ac.cn, lilu@iphy.ac.cn, shenjie@iphy.ac.cn, jphu@iphy.ac.cn

†These authors contributed equally to this work

**The PDF file includes:**

Supplementary Text
Figs. S1 to S20
Tables S1

# Table of content

## 1. Single crystal growth and crystal structure of $K_2Cr_3As_3$

Single crystals of $K_2Cr_3As_3$ were grown with a self-flux of KAs method. First, KAs was prepared by a solid-state reaction. The stoichiometric potassium (solid, 99.95%) and arsenic (lump, 99.99%) were mixed in an Ar-filled glove box. The mixtures were loaded in an alumina crucible and then sealed in an evacuated quartz tube. The sample-loaded quartz tube was heated to 473 K for 10 hours and dwelt for 20 hours. Second, the precursor KAs powders, chromium (powder, 99.99%) and arsenic (lump, 99.99%) were mixed in a molar ratio of 6:1:1, and the mixtures were loaded in an alumina crucible. The alumina crucible (covered with a cap) was jacketed in a Ta tube, and the welded Ta tube was finally sealed in an evacuated quartz tube. The quartz tube was heated up to 1273 K and held for 24 hours in a furnace, followed by cooling down to 923 K at a rate of 2 K/h. Finally, extra flux was removed to obtain single crystals by centrifugation at 923 K. Shiny needle-like crystals were harvested. The synthesized crystals were very reactive in air and easily deteriorate at ambient conditions. These samples were kept in an argon-filled glove box (*40*).

$K_2Cr_3As_3$ crystallizes in a hexagonal lattice with a = 9.9832(9) Å, c = 4.2304(4) Å, and the space group of $K_2Cr_3As_3$ is $P\bar{6}m2$ (No. 187) at room temperature (*17,21,22,24*). The prominent structural unit is the quasi-1D $[(Cr3As3)^{2-}]\infty$ chains along the c-axis, which are arranged triangularly in the ab-plane (*17,21-24,40*). The chains are well separated by $K^+$ ions. Hence, the inter-chains coupling is expected to be much weaker than the intra-chains coupling, exhibiting a strongly 1D structure character (*17,21-24,40*).

## 2. Device fabrication

The $K_2Cr_3As_3$ is sensitive to oxygen and water (*40*), and thus all the microfabrication processes involving exposed crystals were done in a nitrogen-filled glove box. The $K_2Cr_3As$ stripes were first mechanically exfoliated onto the *p*-doped Si substrates with $SiO_2$ layer and spin-coated with PMMA. The samples were transferred and covered by electron-beam deposition of $AlO_x$ films without exposure to air. The SQUID geometries were then patterned using electron-beam lithography, with the subsequent development process also done in the glove box. Afterwards, the $K_2Cr_3As$ stripes were etched by the argon plasma and then deposited by Al leads. After lifting off in the nitrogen-filled glove box, a layer of $AlO_x$ was coated to protect the samples. Now with the protection layer, the samples were taken outside the glove box for wire bonding and measurement. The area of the SQUID is estimated considering the finite penetration length and field distortion effects.

**3. φ-state SQUID as superposition between 0- and π-state SQUID**

The φ-state can be understood as the co-existence of both 0- and π-states of the supercurrent circulating in the whole SQUID. For the conventional 0-state SQUID the total phase in the loop is $\varphi_A - \varphi_B + 2\pi L(I_A - I_B)/\Phi_0 = 2\pi\ \Phi/\Phi_0$, and the supercurrent of the SQUID is simply $I_0(\varphi_A, \varphi_B) = I_A(\varphi_A)+I_B(\varphi_B)$. Here the subscripts A and B denote the two branches of the SQUID. On the other hand, when an additional π-phase is introduced (*4,11*) in the loop, a π-state SQUID is formed with $\varphi_A - \varphi_B + 2\pi L(I_A - I_B)/\Phi_0 + \pi = 2\pi\Phi/\Phi_0$ and $I_\pi(\varphi_A, \varphi_B) = I_A(\varphi_A)+I_B(\varphi_B)$. In JJs with high transmission *t*, the CPR acquires higher harmonics and is usually sufficient to consider the first two harmonics, then both junction A and B have two harmonics: $I_A(\varphi_A) = I_{A1}\sin(\varphi_A) - I_{A2}\sin(2\varphi_A)$ and $I_B(\varphi_B) = I_{B1}\sin(\varphi_B) - I_{B2}\sin(2\varphi_B)$.

When both 0-state and π-state supercurrent co-exist in the SQUID, the supercurrent can be written as: ($p$ and $q$ are the proportions of 0-state and π-state components, respectively, and $p$ and $q$ are real and positive numbers with $p+q = 1$ and $I^*_{A1} = (p-q)I_{A1}$):

$$I_{tot}(\varphi_A,\varphi_B) = pI_0(\varphi_A,\varphi_B) + qI_\pi(\varphi_A,\varphi_B)$$

$$= p[I_{A1}\sin(\varphi_B - 2\pi L(I_A - I_B)/\Phi_0 + 2\pi\Phi/\Phi_0)$$

$$- I_{A2}\sin2(\varphi_B - 2\pi L(I_A - I_B)/\Phi_0 + 2\pi\Phi/\Phi_0)]$$

$$+q[I_{A1}\sin(\varphi_B - 2\pi L(I_A - I_B)/\Phi_0 + 2\pi\Phi/\Phi_0 - \pi)$$

$$- I_{A2}\sin2(\varphi_B - 2\pi L(I_A - I_B)/\Phi_0 + 2\pi\Phi/\Phi_0 - \pi)] + I_B(\varphi_B)$$

$$= (p-q)I_{A1}\sin(\varphi_B - 2\pi L(I_A - I_B)/\Phi_0 + 2\pi\Phi/\Phi_0)$$

$$- I_{A2}\sin2(\varphi_B - 2\pi L(I_A - I_B)/\Phi_0 + 2\pi\Phi/\Phi_0) + I_B(\varphi_B)$$

$$=(p-q)I_{A1}\sin(\varphi_A) - I_{A2}\sin2(\varphi_A) + I_B(\varphi_B)$$

$$= I^*_{A1}\sin(\varphi_A) - I_{A2}\sin(2\varphi_A) + I_B(\varphi_B) \quad \text{(S1)}$$

Therefore, the loop supercurrent with superposition of 0-state and π-state behaves in the similar manner as that having a JJ with the admixture of 0-state and π-state in the loop (equation (1A) in the main text). When the weights between the 0- and π-state are similar, the first harmonic of the junction A is much reduced as $p$-$q$ is much smaller than 1 while the second harmonic is not, thus producing the φ-state with $g = |I_{A2}/I^*_{A1}| > 1/2$.

## 4. Quasi-polar SQUIDs for *p*-wave superconductivity

Figs. S1(A), (B) illustrate the ideal phase-sensitive measurement for detecting *p*-wave superconductivity similar to the theoretical proposal (*11*). As in the main text, we name them as the "polar SQUID" (Fig. S1(A)), which mimics the term "corner SQUID" (*1*) for *d*-wave-YBCO phase-sensitive measurement. When the direction of the coupling is not reversed, we adopt the name "edge SQUID" (Fig. S1(B)), according to the similar setup for *d*-wave-YBCO measurement (*1*).

The theoretical proposal (*11*) considers only the π-phase contribution to the polar SQUID, and only 0-phase contribution to the edge SQUID. However, in realistic devices, both 0- and π-phases are present and we use the device configuration in Fig. S1(E) for most of our devices because of the following reasons:

First, the experimental SQUIDs, even if designed strictly as the theoretical polar SQUID in Fig. S1(A), will correspond to several more realistic configurations such as Figs. S1(C)-(D). One typical situation is the interfaces between *s*-wave and *p*-wave superconductors are not perfectly straight (sketched in Fig. S1(C)) due to the etching direction as proved in Fig. S2(C). Another is the interface is not atomically-smooth and with microfacets (sketched in Fig. S1(D) and shown in Fig. S2(D)), which also appear in other systems such as the YBCO grain boundary with a typical length scale of ~1nm (*8,49*). In such realistic devices, supercurrent flowing through all the possible facets than just along c-axis also provides the admixture of both 0- and π-phases (similar situation was discussed also in d-wave PSM (*1,4,8,49*). We term them as the "quasi-polar SQUIDs". Therefore, they are in principle similar to the design in Fig. S1(E) which is with partially etched facets.

Second, because of non-uniform shape of $K_2Cr_3As_3$ and anisotropic etching, it is very challenging to just etch through $K_2Cr_3As_3$ without further etching the $SiO_2$ substrate underneath, required by the ideal polar SQUID. By taking the TEM, we indeed observe a thin layer of sputtered dust of $SiO_2$ onto the etched surfaces of the $K_2Cr_3As_3$ (Figs. S2(B),(E)) (*59*). Such unintentional $SiO_2$ layer is undesirable for the PSM because: a. Different parity of the order parameters between the *s*-wave and *p*-wave superconductors requires twice the *s*-*p*-*s* reflections in comparison to the single reflection in the conventional *s*-i-*s* Josephson junctions (*61*), that is, the coupling between *s*-wave and *p*-wave pairing is related to $t^4$ ($t$ is the transmission of the interface), instead of $t^2$ for the *s*-*s* Josephson coupling. Therefore, any sub-optimal interfaces will result in a rapid reduction of the *s*-*p* Josephson coupling. b. The weak *s*-*p* Josephson coupling will further reduce or even remove the π-phase contribution, making it insufficient to cancel the first harmonic beyond the ballistic junction limit $g = |I_2/I_1| = 1/2$ and form the φ-state (*60*). Therefore, as shown in details in the zoomed-in figure of the main text Fig. 1(C), the quasi-polar SQUID with partially etched $K_2Cr_3As_3$ can to some extent prevent the side-wall deposition of $SiO_2$ particularly on the c-facet, which is crucial for the *p*-wave PSM if the *p*-wave pairing is along c-axis. Meanwhile, as discussed before, the quasi-polar SQUID with partially etched facets in Fig. S1(E) is in principle the same as Figs. S1(C),(D) in view of the PSM which is still sensitive to the π-phase supercurrent in the 0-π admixture. This justifies our adoption of such configuration (Fig. S1(E) and Fig.1(C)) for most of our devices.

Finally, we emphasize that the third contact ($JJ_C$ in Fig. 1(C)) in the middle of $K_2Cr_3As_3$ is simply used for connecting the quasi-polar SQUID to the external circuit and does not participate in the PSM. Its introduction does not break the continuity of the $K_2Cr_3As_3$ crystal. Also in SM Sec. 12, we explicitly demonstrate the $JJ_C$ is in series with the SQUID because its supercurrent does not oscillate with magnetic field and is well separated with the SQUID's oscillating supercurrent.

We would like to emphasize again that the experimental realization of PSM for *p*-wave superconductor is challenging, because of the possible complex phases in realistic materials and non-ideal situations rising from device nanofabrication. However, with the above reasons, it is justified that the adoption of the partially etched quasi-polar SQUID design in our measurement is a theoretically valid and experimentally viable phase-sensitive test for the π-phase supercurrent of *p*-wave superconductivity. Even supercurrents flowing along both c-axis and a/b-axis are included in this quasi-polar SQUID, it might still be more sensitive to that of c-axis, as will be discussed in the following SM Sec.5.

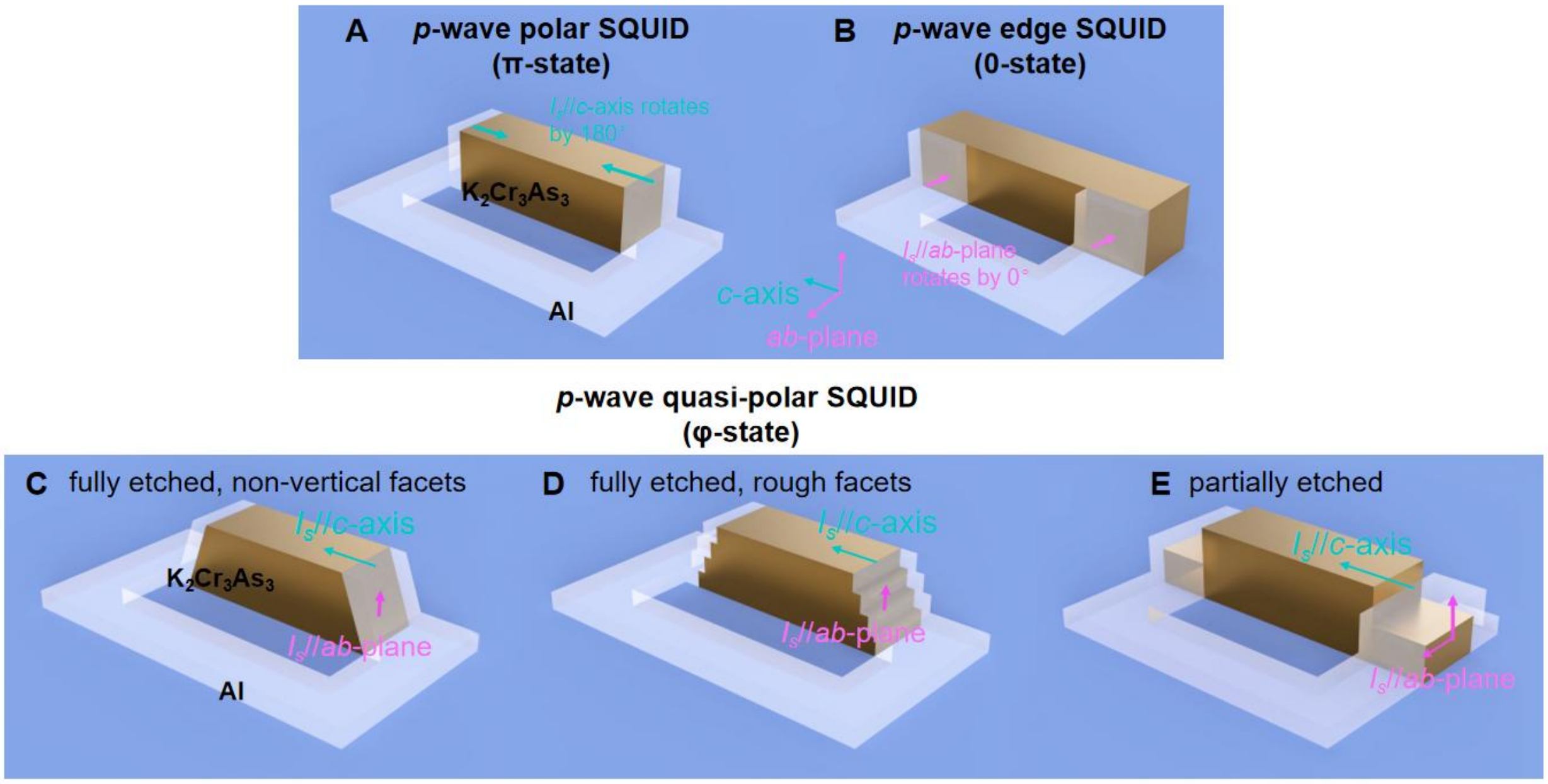

**Fig. S1. Realistic "quasi-polar SQUIDs" in *p*-wave superconductivity PSM. (A)** polar-SQUID for detecting *p*-wave superconductivity in $K_2Cr_3As_3$. The vector of the supercurrent across the Al-$K_2Cr_3As_3$ interfaces rotates by 180° and should introduce a π-phase in the SQUID supercurrent. **(B)** edge SQUID where the vector of the supercurrent across the Al-$K_2Cr_3As_3$ interfaces rotates by 0° and introduces only 0-phase in the supercurrent. **(C)-(E)** *p*-wave quasi-polar SQUIDs with partially (E) or fully etched facets which is non-vertical (C) or rough with microfacets (D). Due to different facets involved in the supercurrent, both 0- and π-phase should be present in such setups. For (C)-(E), the details, such as the $SiO_2$ barrier which will be discussed later, is not concluded here.

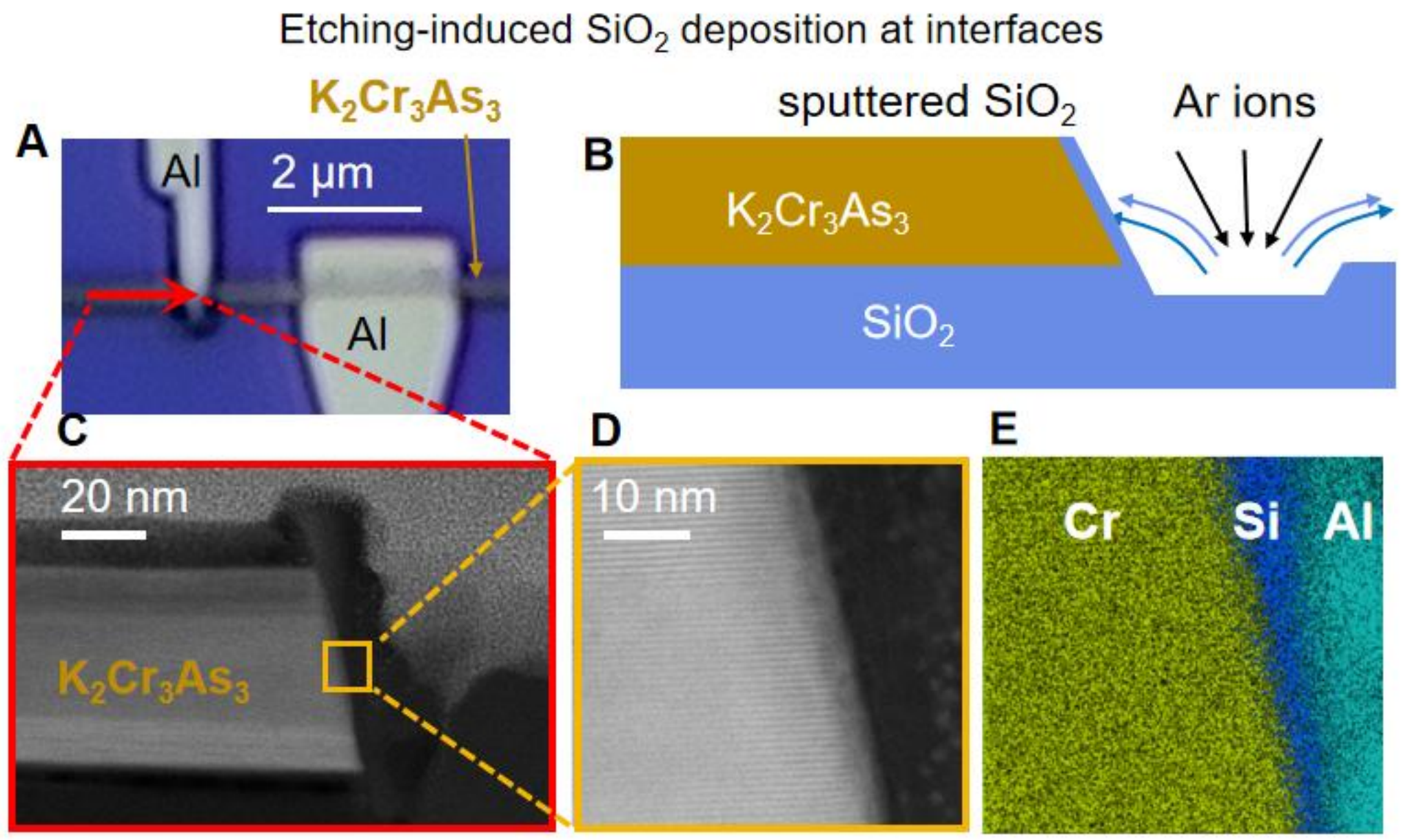

**Fig. S2. Unintentional $SiO_2$ deposition in the fully etched quasi-polar SQUIDs. (A)** optical image of the device showing direction of the TEM cut (red arrow). **(B)** illustration of unintentional $SiO_2$ deposition at the $K_2Cr_3As_3$ interfaces, which is very normal during etching process (*59*). **(C)** TEM image of cross-section in the device showing the slant interface of Al and the fully etched $K_2Cr_3As_3$. **(D)** zoomed-in TEM showing the non-vertical and amorphous interface with microfacets at the order of magnitudes of 1nm. **(E)** EDS analysis showing $SiO_2$ deposition at the fully-etched interface.

## 5. Possible π-phase supercurrent in c-axis

Our data further support that the π-phase supercurrent might be mainly from c-axis (notably, without attempting to exclude that there is also π-phase contribution from a/b-axis). The reasons are as follows:

1. In such configuration, the phase of the supercurrent is less sensitive to *p*-wave pairing along a/b-axis than c-axis: Fig. S3(A) illustrates a single junction of the partially etched quasi-polar SQUID. Following Fig. 1(C), we here call the facets pointed out in by the black arrows as "a/b/c-facets", respectively (Fig. S3(A)). It shows that the supercurrent flowing in the ab-plane indeed rotates by 180° with respect of b-axis (pink arrows) and thus may detect π-phase supercurrent if *p*-wave pairing exists in the b-axis. However, since there are two such junctions in the quasi-polar SQUID, the π-phase contribution at each junction gives a 2π-phase (or 0-phase) supercurrent in the whole SQUID. Therefore, the whole SQUID can only show 0-phase supercurrent, no matter whether the b-axis contains *p*-wave superconductivity or not. Besides, the interfaces between the Al-"b-facets" include an unintentionally sputtered $SiO_2$ layer (yellow arrow in Fig. S3(A)) even in the partially-etched case, which also reduces the Josephson coupling in the b-axis direction. In addition, this device configuration does not rotate by 180° with respect of a-axis, and thus is not related to the *p*-wave superconducting pairing along a-axis. Therefore, our partially-etched quasi-polar SQUID is insensitive to *p*-wave superconductivity in a/b-axis. On the other hand, light blue arrows in Fig. S3(A) shows a non-rotating, stable n-vector for the supercurrent flowing in the c-axis. Such vector rotates by 180° throughout the two junctions in the SQUID (Fig. 1(C)) thus providing exactly the condition required by *p*-wave PSM in the c-axis (*11*). Therefore, our design is more sensitive to π-phase supercurrent originating from the *p*-wave superconductivity in c-axis.
2. Although the supercurrent may flow across different interfaces (Al-a/b/c axis), their tunability by the magnetic field is different. As shown by Fig. 3(A) and Fig. S3(B), due to the Meissner effect, the field is distorted in the ab-plane. Therefore, the $I_s$ parallel to the ab-plane (pink arrows) is not much exposed to the field and thus cannot be efficiently controlled by it. On the contrary, the supercurrent flowing across the Al-"c-facet" interface (light blue arrows in Fig. S3(A)) is less affected by the screening and could be effectively modulated by the magnet field. It is clearly shown that in Fig. 3 the π-phase supercurrent component is highly tunable by the magnetic field, and thus should correspond mainly to the supercurrent flowing across in c-axis, similar to the modulated $I_c$ oscillation pattern of SQUID by flux pinning through YBCO grain boundary junctions (*8,49*).
3. We also measured some devices where $K_2Cr_3As_3$ is completely etched and the JJs only connect to c-axis (as Figs. S1(C),(D)). As shown later in Device E in Figs. S11(E),(F), we observe there the π-periodic oscillation characteristic to the φ-junction, thus further demonstrating the presence of π-phase components in supercurrent flowing in c-axis. The weak π-periodic oscillation occurred at higher field is also consistent with the reduced *s-p* coupling by unintentional $SiO_2$ deposition at the interface in the fully-etched devices.
4. In Fig. S4, we also fabricate different SQUIDs on the same $K_2Cr_3As_3$, which are one edge SQUID (R2) and two quasi-polar SQUIDs along c-axis (R1) and a/b-axis (R3) respectively. We only observe the φ-state with π-periodic oscillation in R1 (Fig. S4(H)) around zero field, which means the π-phase from c-axis is more predominating than that from the a/b-axis. We note that here we do not attempt to exclude the possible π-phase contribution from other two SQUIDs in other situation, for example, at higher magnetic field or lower temperature. Our data simply indicate that the π-phase along c-axis is much stronger than other two cases.

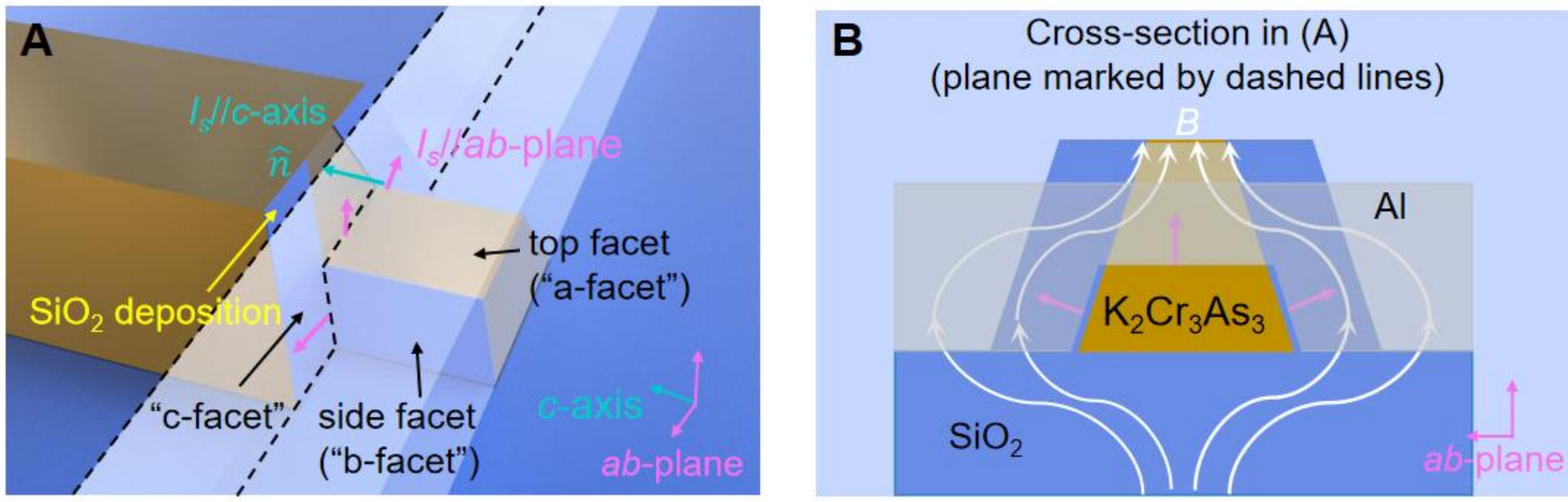

**Fig. S3. Models illustrating the field distortion and non-ideal interfaces at a Al-$K_2Cr_3As_3$ junction. (A)** 3D model zooming in to a single junction area of the partially-etched quasi-polar SQUID. $SiO_2$ also partially deposited (yellow arrow) on the interfaces even in the partially-etched quasi-polar SQUIDs, though not entirely as in the fully-etched case in Fig. S2(B). Supercurrent at the Al-$K_2Cr_3As_3$ interfaces has the component in c-axis with a stable n-vector (light blue arrows) and components in the ab-plane with rotating n-vector (pink arrows). **(B)** 2D cross-section at the plane marked by the dashed lines in (A). The distorted magnetic field lines penetrates between c-facet of $K_2Cr_3As_3$ and Al film in the penetration length, but wraps around the remaining $K_2Cr_3As_3$ section. $I_s$//ab-plane (pink arrows) is not efficiently exposed to $B$ due to Meissner effect. On the contrary, $I_s$//c-axis (blue arrows in (A)) could be modulated by $B$.

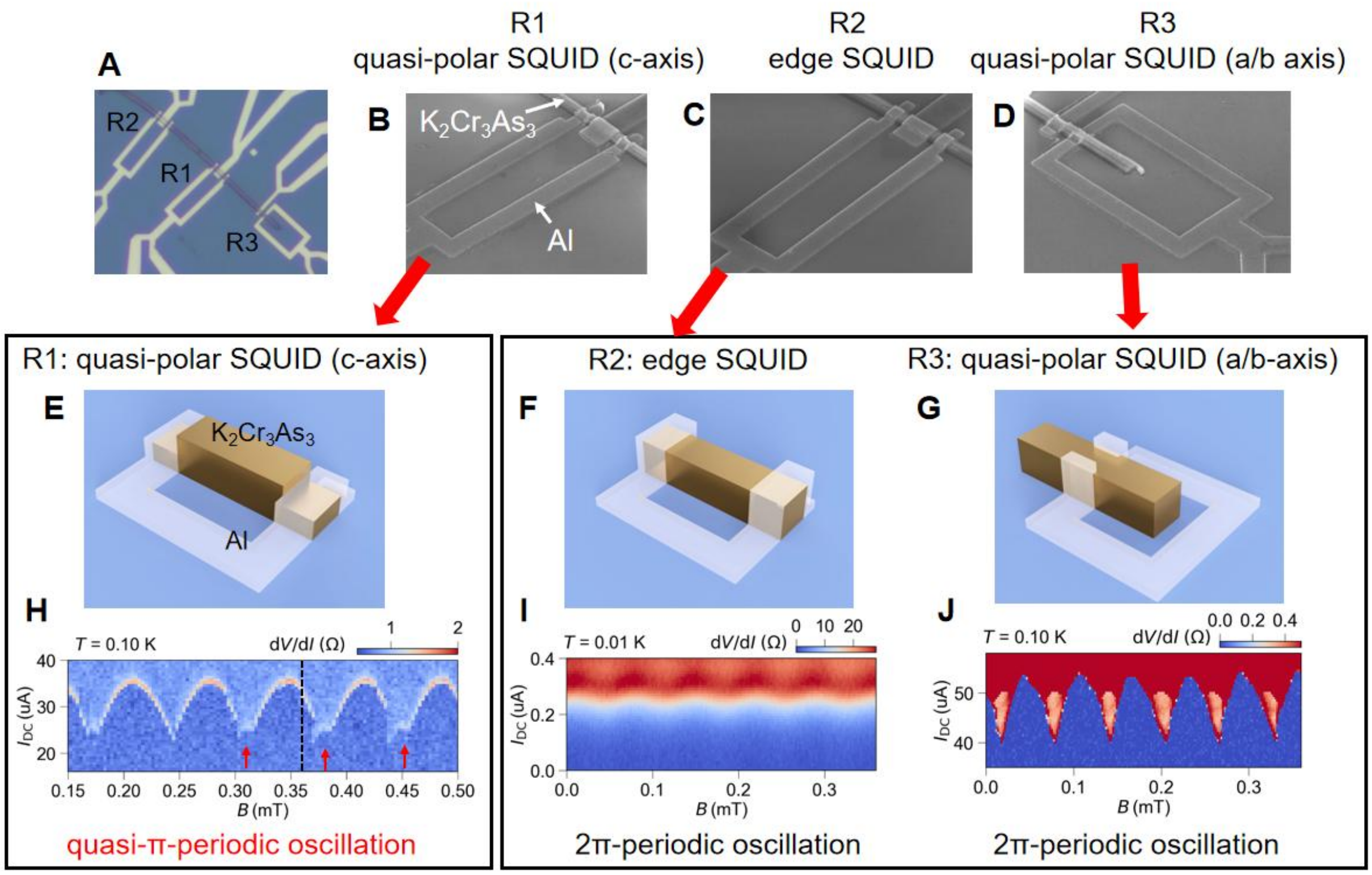

**Fig. S4. Comparison between quasi-polar SQUID in the c-axis (R1) and a/b-axis (R3) and the edge SQUID (R2)**. **(A)** optical image showing three devices fabricated on the same $K_2Cr_3As_3$ crystal. **(B)-(D)**, SEM images of the three devices. In R2 the $K_2Cr_3As_3$ crystal is very lightly etched so that the surface area in the "c-facets" (defined in Fig. S3(A)) is much smaller than that of R1 which is etched heavily. In R3, the etching recipe is the same as that of R1, but along "b-facets" instead of "c-facets". **(E)**,**(H)** schematic of R1 and the measured $I_c(B)$ oscillation of the supercurrent showing quasi-π-periodic feature (red arrows) and thus high π-phase component. $T$ = 100 mK. The dashed vertical line marks the same upper $B$ bound as to (I) and (J). **(F)**,**(I)**, schematic of R2 and the measured $I_c(B)$ showing only the normal 2π-periodic oscillation. $T$ = 10 mK. **(G)**,**(J)** schematic of R3 and $I_c(B)$ showing only the 2π-periodic oscillation. $T$ = 100 mK. The details of the devices, such as the $SiO_2$ barrier at the interface, are not included in the schematics in (E),(F),(G).

## 6. Fast counter measurement of $I_c$(B) in Figure 3

The many oscillations over wide range of $B$ in Fig. 3 of the main text are obtained by the fast counter measurement technique commonly adopted in SQUID characterization (*50,62*). In such method, $I_{DC}$ is ramped linearly and continuously from 0 to above $I_c$, and the dc voltage of the SQUID is monitored. When $I_{DC} = I_c$, the voltage will jump from 0 (the superconducting state) to the finite value (the normal state) which triggers a counting event of a digital counter. The time interval between $I_{DC} = 0$ and $I_{DC} = I_c$ obtained by the counter thus provides a fast measurement of $I_c$ over a wide range of $B$. The pulse frequency $f$ is much smaller than the cut-off frequency of the lock-in amplifier so that $R$ is not distorted.

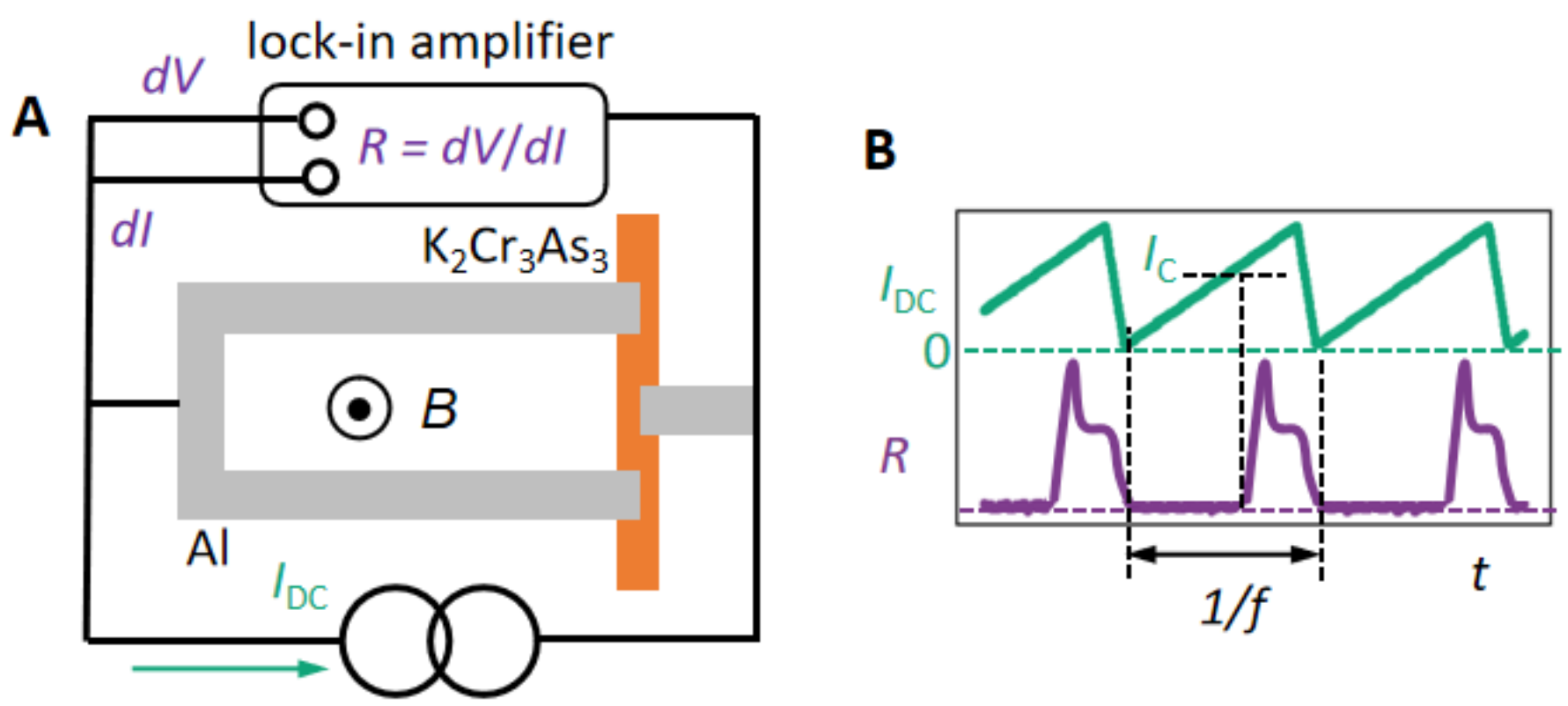

**Fig. S5. Schematic showing fast-pulse measurement setup. (A)** Setup involving a lock-in amplifier and a fast pulse generator providing the triangular wave of $I_{DC}(t)$ (green waveform in (B)). **(B)** Time-domain signals for $I_{DC}$ and $R$ with frequency $f$. $I_c$ is determined at $t$ where $R$ becomes non-zero.

**7. Extended data for Figure 3**

Fig. S6(A) shows the similar data as Fig. 3(A) of the main text measured over a wide range of $B$ at $T$ = 0.3 K, again showing a transition where the π-periodic oscillation is the clearest. Fig. S6(B) also shows the wide sweeps at several temperatures.

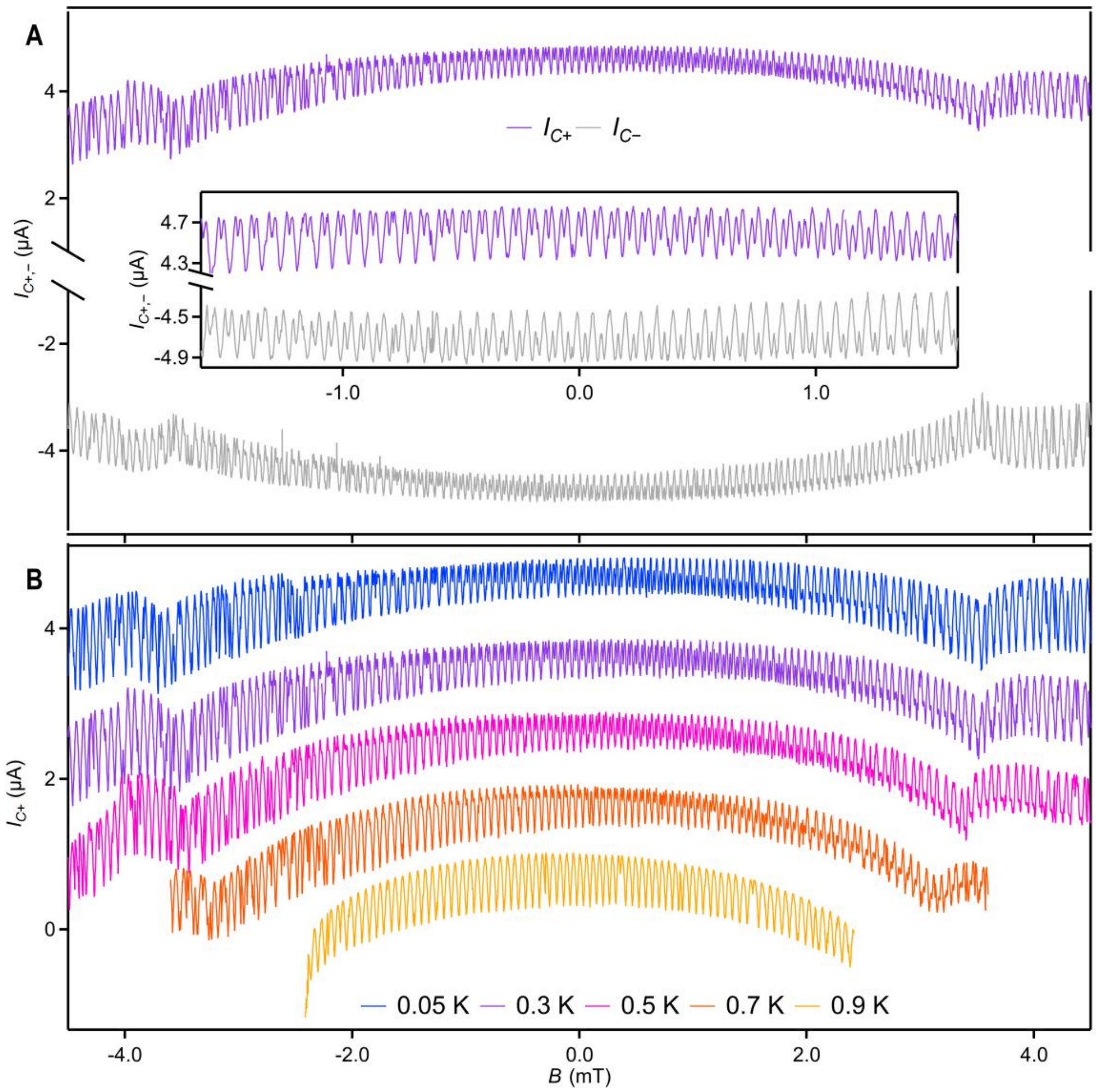

**Fig. S6. Extended data for Fig. 3(A) with more temperature points. (A)** Positive and negative critical current $I_{c+,-}(B)$ at $T$ = 0.3 K, again showing π-to-2π-periodic transition, similar to the 0.1 K data in Fig. 3(A) of the main text. Inset: the zoomed-in plot across the transition region. **(B)** $I_{c+}(B)$ for all other measured temperatures $T$ = 0.05, 0.3, 0.5, 0.7 and 0.9 K. The curves in (B) are offset in y-axis for clarity.

## 8. Fourier analysis of $I_c$(B)

The measured $I_c(B)$ in the main Fig. 3 is reproduced in Fig. S7, and the amplitude of the Fourier spectrum is numerically calculated by the fast Fourier transform (FFT) algorithm. The result from the direct calculation is plotted in Fig. S7(B) (black dots). The amplitude is enhanced at $n = \Delta B/B = 1$ and 2, indicating the strong first and second harmonic of $I_c(B)$ shown in Fig. S7(A). Adopting the common practices for Fourier analysis (*63*), duplicating the data by several periods further sharpens the harmonic peaks (blue dot). Appending a string of zeros ("zero-padding") to the duplicated data also enhances the frequency resolution of the FFT (red dot). In the main figures, we consistently use the method of period duplication and zero padding to calculate the Fourier spectrum and to accurately extract the first and second harmonics.

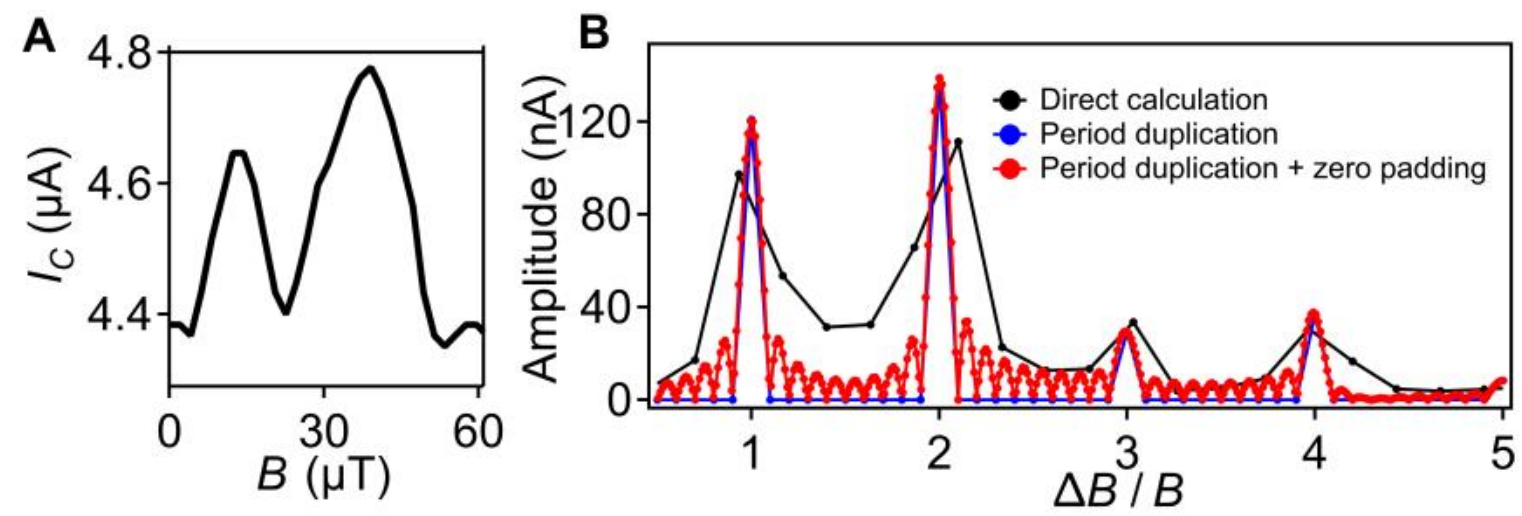

**Fig. S7. Fourier analysis for $I_c(B)$. (A)** $I_c(B)$ data. **(B)** FFT amplitude spectrum of the data in (A), calculated in three methods: 1. direct calculation (black), 2. period duplication (blue), 3. period duplication and zero padding (red, also the one used in the main text). $\Delta B$ is the length of the *x*-axis in (A). $\Delta B/B$ = 1, 2, 3,... correspond to the first, second, third harmonics, etc (*63*).

**9. Fitting of $I_c$(B)**

Figs. S8(A), (B) show the original 2D plots used to extract $I_c(B)$ in Fig. 2(E). As described in the main text equation (1), the SQUID in the $\varphi$-JJ state can be modeled as the two Josephson junctions in parallel (noted as $JJ_A$ and $JJ_B$), each with the CPR $I(\varphi) = I_1\sin(\varphi) - I_2\sin(2\varphi)$. While generally $I_1$ can be both positive and negative, we restrict their values to positive in the fitting since $g = |I_2/I_1|$ is of the central importance in this work. Since the measured averaged $I_c$ (~ 4 μA) is around ten times higher than its flux variation (~ 0.4 μA), it is justified to simplify the CPR of the strong junction (say $JJ_B$) as purely sinusoidal, without affecting the shape of $I_c(B)$. The inductance of the SQUID is calibrated using the data at high temperature $T$ = 1.0 and 1.1 K, where we neglect the second harmonic. This results in $L$ = 88 pH (or $\beta = 2\pi L I_c/\Phi_0 \approx 0.37$ for $I_c \approx 1.33$ μA at 1.1 K). Such value is higher than expected for Al contacts (*64*) but possibly originates from the kinetic inductance (*65*) of $K_2Cr_3As_3$.

By fixing $L$, the extracted $I_c(B)$ in Figs. S8(C) can be fitted by the equation (2) in the main text. In order to increase the accuracy of the fitting, we introduce two fitting parameters: a relative phase shift $\Delta\varphi_2$ between the first and the second harmonics and an extra phase $\Delta\varphi_S$ in the SQUID loop, which are present possibly due to the spontaneous current (*11,33-35,44-46,66*). Therefore, the model for our system can be written as (*1,5,36,67*):

$$I_c(\Phi) = \max_{\varphi_A}\{I_A(\varphi_A) + I_B(\varphi_B)\} \quad \text{(S2A)}$$

$$I_A(\varphi_A) = I_{A1}\sin(\varphi_A) - I_{A2}\sin(2\varphi_A + \Delta\varphi_2) \quad \text{(S2B)}$$

$$I_B(\varphi_B) = I_{B1}\sin(\varphi_B) \quad \text{(S3C)}$$

$$\varphi_A - \varphi_B + 2\pi L[I_A(\varphi_A) - I_B(\varphi_B)]/\Phi_0 = 2\pi\Phi/\Phi_0 + \Delta\varphi_S \quad \text{(S2D)}$$

where $L$ is the inductance in the SQUID loop.

Using equation (S2), we have achieved reasonably good agreement with the data for all temperature points, as demonstrated by overlapping the fitted $I_c(B)$ (yellow) on the 2D plots (Figs. S8(A),(B)). The fitting parameters $I_{A1}$ and $I_{A2}$ are reproduced from Fig. 2(E) in Figs. S8(D). We emphasize that the other fitting parameters $\Delta\varphi_2$ and $\Delta\varphi_S$ (Figs. S8(E)), introduced to increase the accuracy of the fitting, do not affect the magnitudes of the first and second harmonics of the JJs. Indeed, as demonstrated in Fig. 2(F), the FFT, which is insensitive to $\Delta\varphi_2$ and $\Delta\varphi_S$ but directly captures the amplitudes of the first and second harmonics, still gives the almost identical results to the fitting. Therefore, it demonstrates beyond doubt that at low temperatures $g > 1/2$ and the SQUID is in the $\varphi$-state.

Similarly, Figs. S9(A)-(E) reproduce all the segments of $I_c(B)$ used in Fig. 3(D)-(F) taken at the fixed $T$ = 0.1 K but different $B$. By fitting each segment to equation (S2), we obtain the fitting parameters shown in Fig. 9(F) (same as Fig. 3(E)) and 9(G). Again, although $\Delta\varphi_2$ (Fig. 9(G)) is needed to increase the accuracy of the fitting, its inclusion does not alter the harmonic amplitudes $I_{A1}$ and $I_{A2}$ which are almost identical to the FFT in Fig. 3(D).

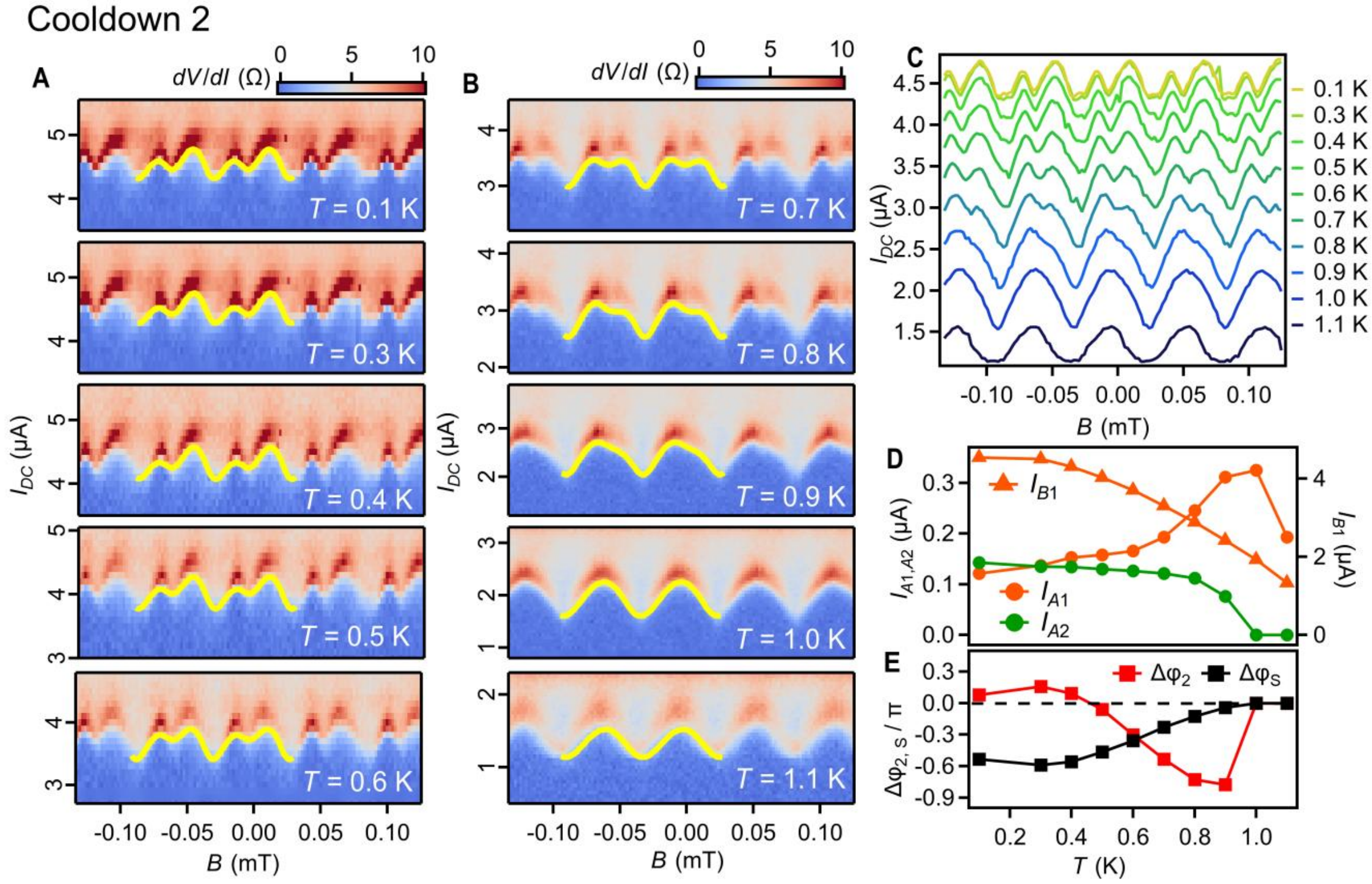

**Fig. S8. Flux-dependent critical current for Cooldown 2 in Fig. 2. (A)-(B),** $dV/dI(I_{DC}, B)$ at $T$ = 0.1, 0.3, 0.4, 0.5, 0.6, 0.7, 0.8, 0.9, 1.0 and 1.1 K are the original 2D plots used for harmonics analysis in Fig. 2(D)-(F). The fitting results of $I_c(B)$ are overlapped for all the figures. **(C)** Extracted $I_c(B)$ at each $T$. **(D)** Fitting parameters (the two harmonics of $JJ_A$ ($I_{A1,2}$) and the first harmonic of $JJ_B$ ($I_{B1}$)) versus $T$. The data is the same as Fig. 2(E). **(E)** Fitting parameters (the relative phase shift between the two harmonics of $JJ_A$ ($\Delta\varphi_2$) and the extra phase shift between $JJ_A$ and $JJ_B$ possibly due to the self-generated flux ($\Delta\varphi_S$)) versus $T$.

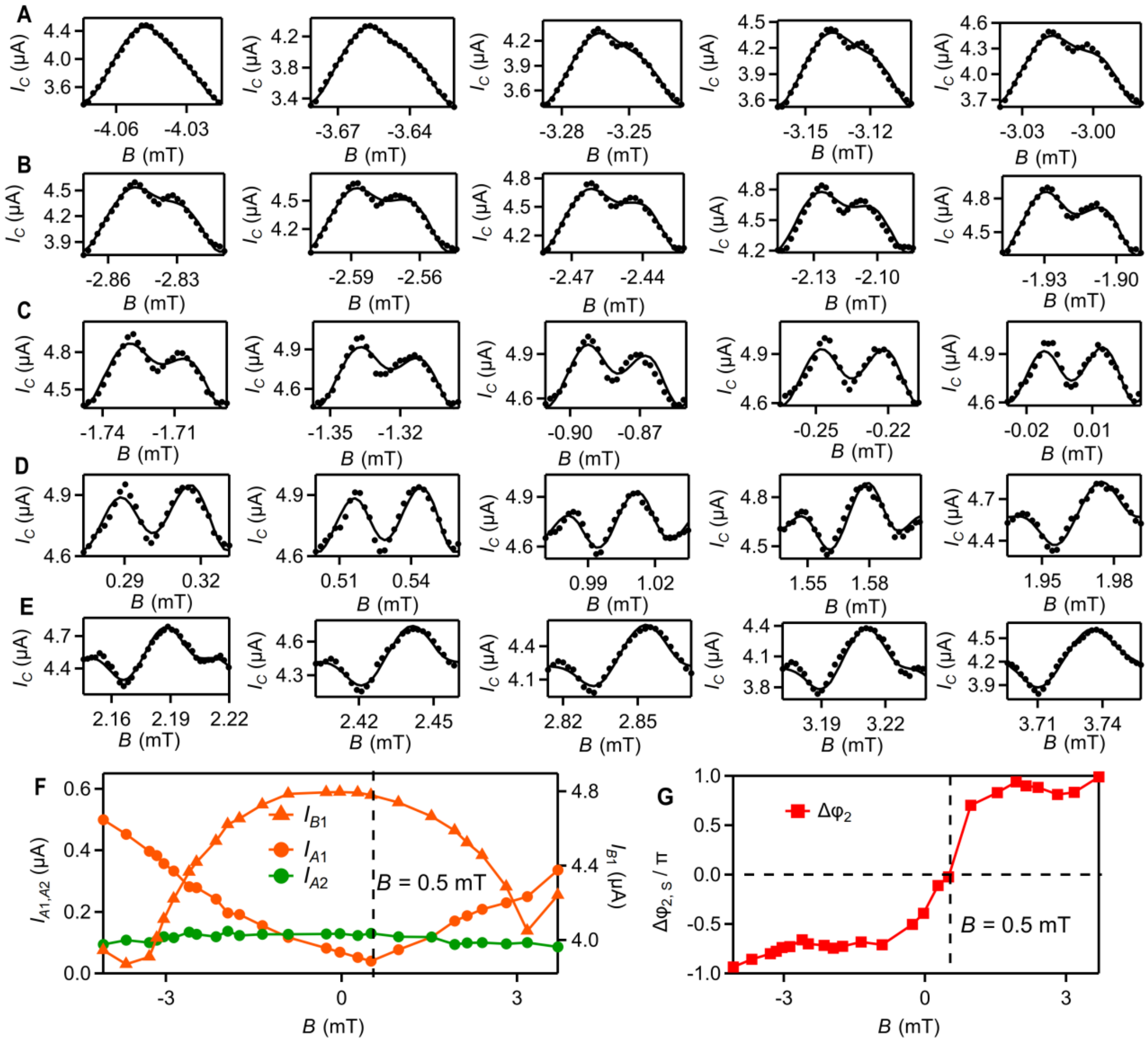

**Fig. S9. Extended data for Fig. 3. (A)-(E)** Extracted $I_c(B)$ at each $T$. All the measured $I_c(B)$ segments at different $B$ (dots) with the fitting curves (solid lines) used for Fig. 3(E) in the main text. **(F)** The fitting parameters $I_{A1,2}$ and $I_{B1}$ versus $B$ reproduced from Fig. 3(E). Each point corresponds to a $I_c(B)$ panel in (A)-(E). $g = I_{A2}/I_{A1}$ is maximal around $B$ = 0.5 mT. **(G)** $\Delta\varphi_2$ versus $B$, $\Delta\varphi_2(B)$ is roughly an odd function with respect to $B$ = 0.5 mT.

## 10. More devices with and without quasi-π-periodic features

Devices with quasi-π-periodic features

We first show other similarly fabricated devices which also exhibit the quasi-π-periodic structure in the phase-sensitive measurement. The optical image of the SQUID device (Device B) with similar geometry is included in Fig. S10(A). The wider range $I_c(B)$ is plotted in Fig. S10(B), where the quasi-π-periodic features in the SQUID oscillations emerge in the positive $I_{DC}$ for $B$ ~ 5 mT. Similar to Fig. 3, the oscillation pattern is always point symmetric with respect to zero field and such quasi-π-periodic oscillations also appear in the negative $I_{DC}$ for $B$ ~ -5 mT (Fig. S10(C)). The typical segments of $I_c(B)$ (positive current for $B$ ~ 5 mT and negative current for $B$ ~ -5 mT) can also be characterized by FFT, giving g = $F_2/F_1$ = 0.72 > 0.5. Similarly, Device C also shows clear quasi-π-periodic oscillations for both negative and positive $I_{DC}$ near zero field (Figs. S11(A), (B)). Device D where $I_c$ of the SQUID is higher than that of the contact junction $JJ_C$ outside the SQUID, the quasi-π-periodic oscillation is also observed ~ 0.7 mT (Figs. S11(C), (D)). Finally, Device E-G exhibits weaker π-periodic oscillations also at finite fields (Figs. S11(E)-(J)). Devices A-G thus demonstrate the reproducibility of the φ-state in Al-$K_2Cr_3As_3$ quasi-polar SQUIDs.

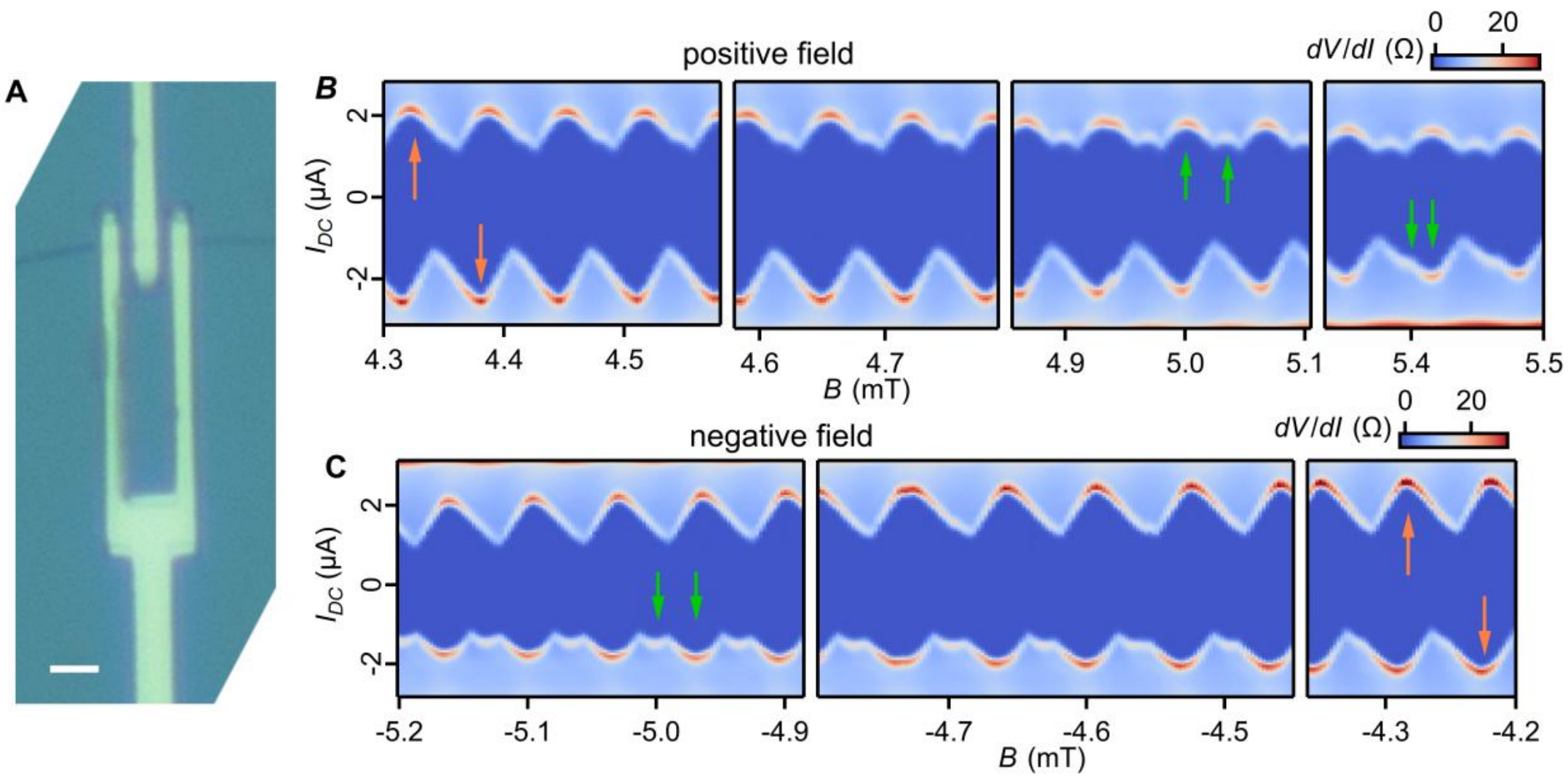

**Fig. S10. Device B showing 2π- to quasi-π-periodic oscillations. (A)** Optical image. Scale bar: 2 μm. **(B)** d*V*/d*I*($I_{DC}$, *B*) measured at *T* = 96 mK, showing a clear transition from the 2π- (orange arrows) to quasi-π-periodic (green arrows) oscillations in the positive $I_{DC}$ for positive fields. **(C)** Same as **(B)** showing 2π- to quasi-π-periodic transition in the negative $I_{DC}$ for the negative fields. The FFT at the quasi-π-periodic oscillations (green arrows) gives $g = F_2/F_1 = 0.72 > 1/2$.

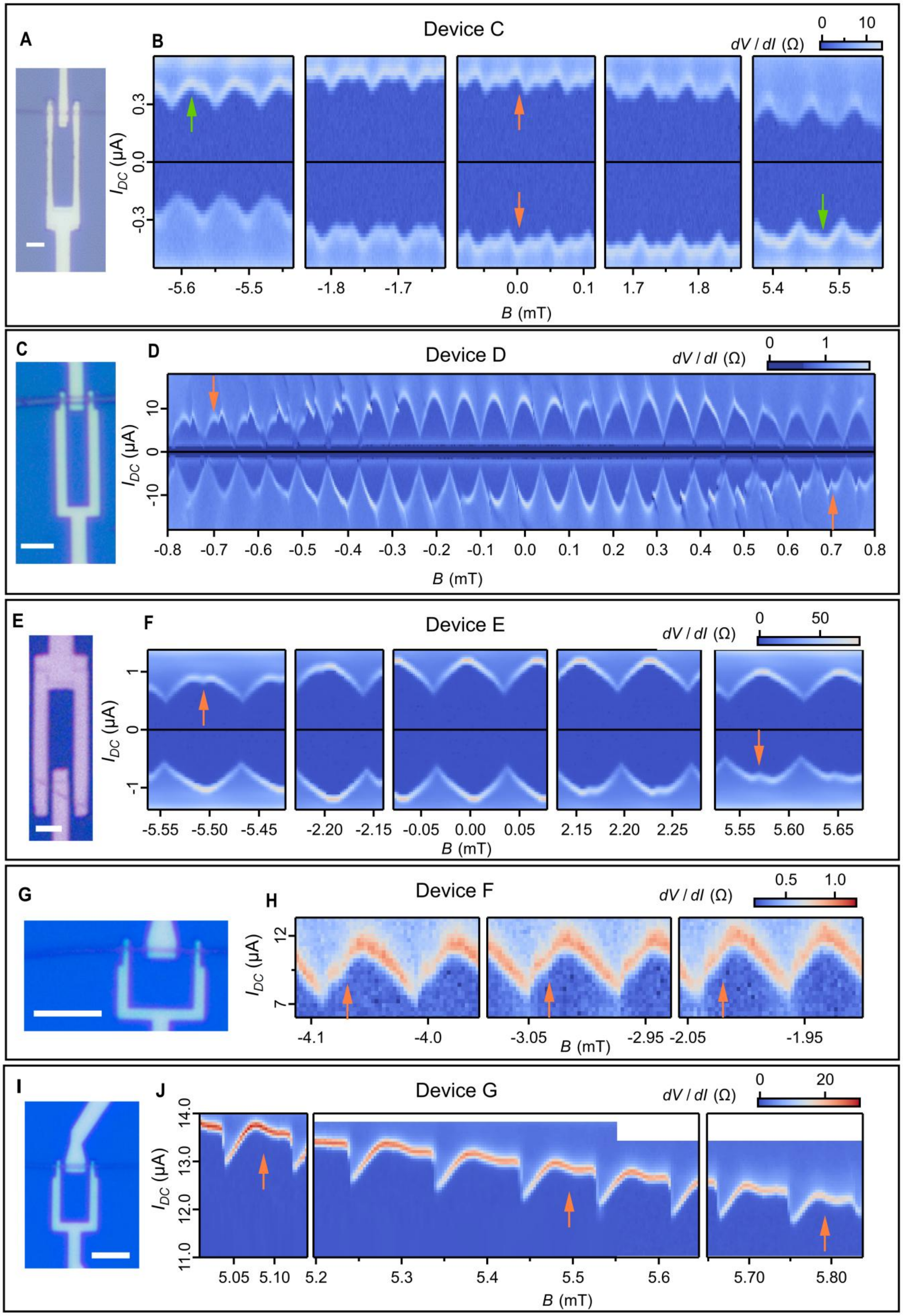

A
B
Device C
dV / dI (Ω)
0 10
I_DC (μA)
0.3 0.0 -0.3
-5.6 -5.5 -1.8 -1.7 0.0 0.1 1.7 1.8 5.4 5.5
B (mT)
C
D
Device D
dV / dI (Ω)
0 1
10 0 -10
-0.8 -0.7 -0.6 -0.5 -0.4 -0.3 -0.2 -0.1 0.0 0.1 0.2 0.3 0.4 0.5 0.6 0.7 0.8
E
F
Device E
dV / dI (Ω)
0 50
1 0 -1
-5.55 -5.50 -5.45 -2.20 -2.15 -0.05 0.00 0.05 2.15 2.20 2.25 5.55 5.60 5.65
G
H
Device F
dV / dI (Ω)
0.5 1.0
12 7
-4.1 -4.0 -3.05 -2.95 -2.05 -1.95
I
J
Device G
dV / dI (Ω)
0 20
14.0 13.0 12.0 11.0
5.05 5.10 5.2 5.3 5.4 5.5 5.6 5.70 5.80

**Fig. S11: Devices C-G showing quasi-π-periodic oscillations. Device C: (A)** Optical image. Scale bar: 2 µm. **(B)** d$V$/d$I$($I_{DC}$, $B$) measured at $T$ = 10 mK, showing a clear quasi-π-periodic oscillations (orange arrows) near zero field. $I_C$($B$) patterns at higher fields obey the Onsager relation (green arrows). **Device D: (C)** Optical image. Scale bar: 3 µm. **(D)** d$V$/d$I$($I_{DC}$, $B$) measured at $T$ = 1.15 K, showing a weak π-periodic oscillations (orange arrows) above ~ 0.7 mT. $I_c$ of the SQUID is higher than that of $JJ_C$. **Device E: (E)** Optical image. Scale bar: 2 µm. **(F)** d$V$/d$I$($I_{DC}$, $B$) measured at $T$ = 10 mK, showing a weak π-periodic oscillations (orange arrows) above ~ 5 mT. **Device F: (G)** Optical image. Scale bar: 5 µm. **(H)** d$V$/d$I$($I_{DC}$, $B$) measured at $T$ = 10 mK, showing a weak π-periodic oscillations (orange arrows) more negative than ~ -2 mT. $I_c$ of the SQUID is higher than that of $JJ_C$. **Device G: (I)** Optical image. Scale bar: 5 µm. **(J)** d$V$/d$I$($I_{DC}$, $B$) measured at $T$ = 10 mK, showing a weak π-periodic oscillations (orange arrows) higher than ~ 5 mT.

Devices without quasi-π-periodic features

Meanwhile, for other devices fabricated via nominally the same process, no quasi-π-periodic oscillations are observed. Fig. S12(A) shows similar device geometry (Device H1). However, for wide range of field, no quasi-π-periodic oscillation is visible in either direction of $I_{DC}$ (Figs. S12(B),(C)).

Several other devices (H2-H5, not shown) with different loop areas and junction areas also show 2π-periodic oscillations up to the critical field of Al. We note that H1-H5 were not measured immediately after fabrication, as Devices A-G were. Indeed, a typical EDS measurement shown in Fig. S13 showing that the element K is partially lost over time, even all the devices are covered by a protection layer of $AlO_x$ and are stored in the glovebox (but still $K_2Cr_3As_3$, instead of degrading into other materials, e.g. $KCr_3As_3$). Therefore, the conventional SQUID oscillation may be due to the gradual degradation of the stripe. Meanwhile, due to the different crystal dimensions, anisotropy etching, and the unintentional deposition of the $SiO_2$ and so on, the devices showing quasi-π-periodic oscillations or not (and thus $\varphi$-JJ state or not) might be due to the interface details, as well as different chemical potential, and thus the different coupling between the crystal and the *s*/*p*-wave superconductors (*17*,*23*,*25*).

Figs. S12(D), (E) also plot the $R = \mathrm{d}V/\mathrm{d}I$ versus $T$ for Devices A-G and H1-H5 respectively, showing no clear correlation between the sharpness of the superconducting transition for $K_2Cr_3As_3$ and the occurence of quasi-π-periodic oscillations. Table S1 lists the average critical current $\overline{I_c}$ and the half of the oscillation amplitude $\Delta I_c$. $\eta = \Delta I_c/\overline{I_c}$ thus describes the symmetry of the SQUID (*68*).

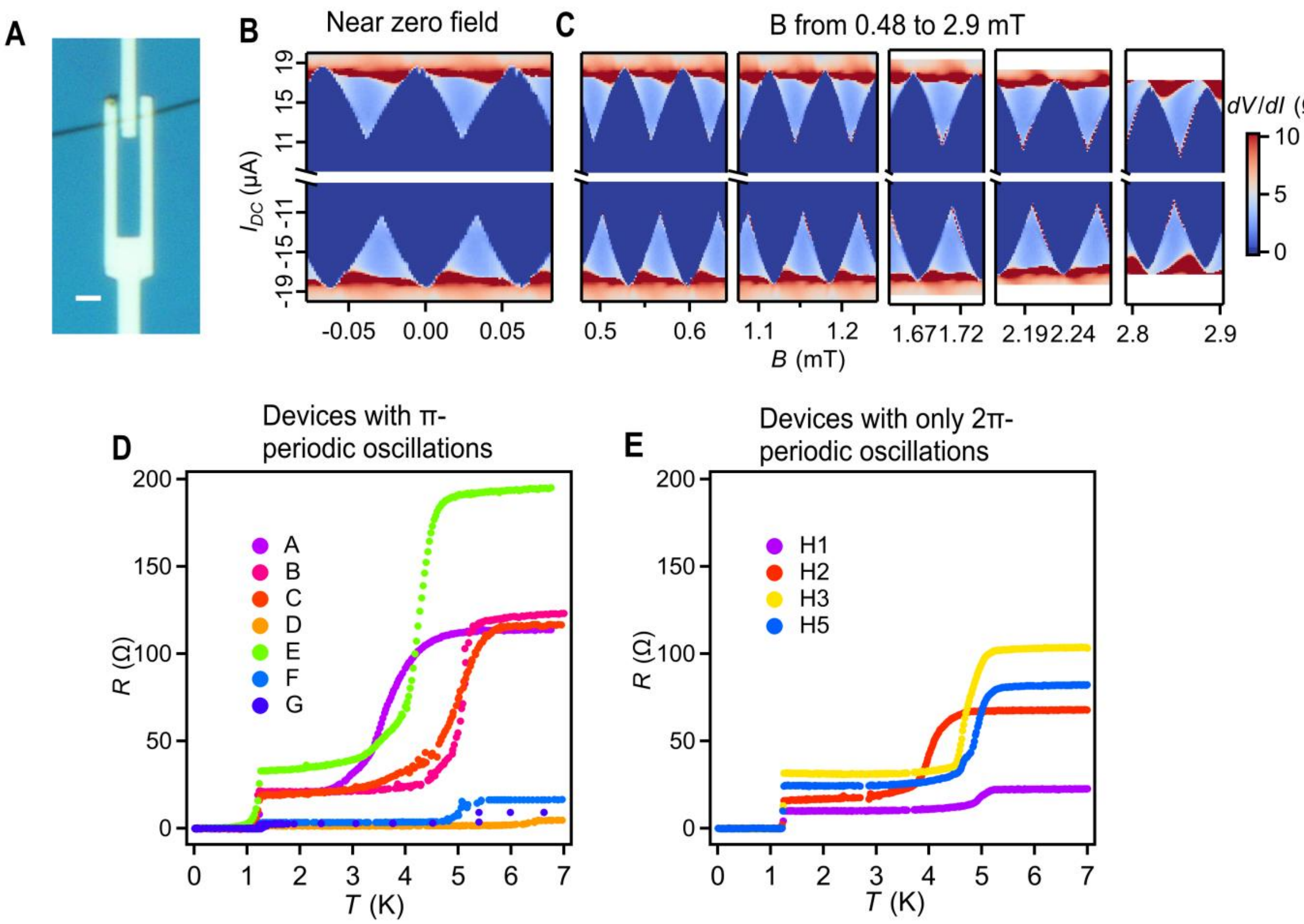

**Fig. S12. Device H1 showing only 2π-periodic oscillations and *R*-*T* curves for all devices. (A)** Optical image of the SQUID. Scale bar: 2 μm. **(B)** d*V*/d*I* versus $I_{DC}$ and *B* near zero field. The critical current $I_c$ shows only 2π-periodic oscillations. **(C)** d*V*/d*I* versus $I_{DC}$ and *B* for higher field. *T* = 93 mK. **(D, E)** *R* = d*V*/d*I* versus *T* for Devices A-G and H1-H5, respectively**.** Devices A-G and Devices H1-H5 showing different sharpness at the superconducting transition temperature for $K_2Cr_3As_3$ without clear correlation with the occurrence of quasi-π-periodic oscillations. The data for Devices A and B are the same as Fig. 1(E).

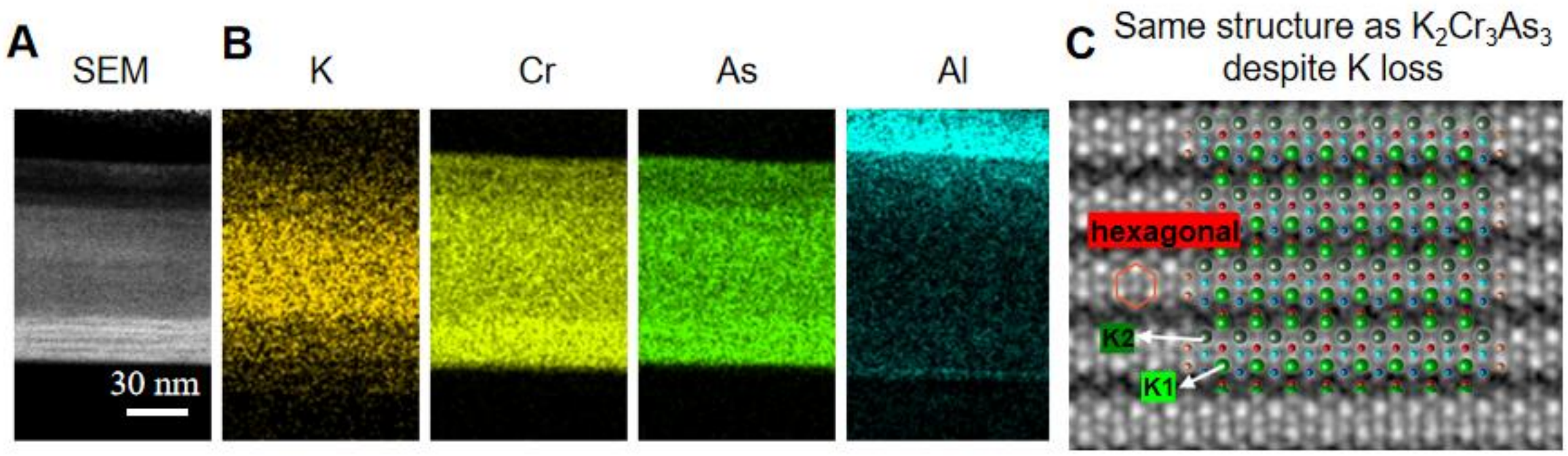

**Fig. S13. EDS analysis of a section of stripe without π-periodic oscillations.** Scale bar: 30 nm. **(A)** The first image is the SEM taken simultaneously. **(B)** The stripe is covered with a protection layer of $AlO_x$ (orange). Contrary to the clear-cut boundaries of Cr (bright yellow) and As (green) which coincide with the SEM image, K (dark yellow) is visibly lost possibly due to aging effect. **(C)** TEM showing that the hexagonal lattice of $K_2Cr_3As_3$ is unchanged despite K loss (not degrading into other materials, e.g. $KCr_3As_3$).

| Devices | $\overline{I_c}$ (μA) | $\Delta I_c$ (μA) | $\eta$ (%) |
|---|---|---|---|
| A | 4.5 | 0.3 | 6.7 |
| B | 1.4 | 0.2 | 14.3 |
| C | 0.38 | 0.048 | 12.5 |
| D | 8.0 | 4.5 | 56.2 |
| E | 0.76 | 0.27 | 35.5 |
| F | 9.29 | 1.77 | 19.1 |
| G | 0.33 | 12.04 | 2.8 |
| H1 | 14.9 | 3.5 | 23.5 |
| H2 | 3.1 | 0.9 | 29.0 |
| H3 | 1.2 | 0.8 | 66.7 |
| H4 | 1.9 | 1.0 | 52.6 |
| H5 | 3.4 | 1.1 | 32.3 |

**Table S1. Summary of Summary of $\overline{I_c}$, $\Delta I_c$ and $\eta = \Delta I_c/\overline{I_c}$ for all devices.** Devices A-G show quasi-π-periodic oscillations. Devices H1-5 show only 2π-periodic oscillations.

## 11. Ruling out residual ferromagnetism in Al-$K_2Cr_3As_3$ SQUIDs

It is possible that the observed φ-state is originated from the residual ferromagnetism of the sample instead of the *p*-wave superconductivity (*33,36*). Josephson devices with ferromagnetic components possess several general properties. First, they will show hysteretic behavior when $B$ is stepped up or down (*69*). Furthermore, different from the classical Fraunhofer pattern with $I_C(B)$ = $I_C(-B)$, JJs may in some occasions display asymmetry between $I_C(B)$ and $I_C(-B)$, possibly due to mixture between conventional and unconventional superconductivity phases (*1*), ferromagnetism (*70-72*), and non-uniform supercurrent distribution or vortices (*73,74*). As illustrated in Fig. S14(A), in the case of ferromagnetism, the asymmetric $I_C(B)$ will show a mirror-symmetric pattern with respect to $B = 0$ when $B$ is stepped up or down across the coercive field (*70-72*).

By checking these properties, we rule out ferromagnetism as the cause for the quasi-π-periodic oscillation of $I_c(B)$ in our devices, for the following reasons: (i) As shown in Fig. S14(B) when the field range is kept within ~0.3 mT, the detailed quasi-π-periodic SQUID oscillation is identical with no hysteresis when $B$ is stepped up or down. Apart from several glitches in the fast SQUID oscillations due to vortices, no hysteresis appears in the Fraunhofer envelop when we increase the range of sweep to 2.5 mT (Fig. S14(C)). (ii) $I_c(B)$ in Fig. S14(C) is slightly asymmetric. However, no mirror symmetry explained before (*70-72*) is present between the patterns with up-stepping and down-stepping $B$. (iii) It remains possible that the ferromagnetic components have the coercive fields much higher than 2.5 mT. To exclude such possibility, we first apply the magnetic field of +12 T to another similar device Devices C and then measure $I_c(B)$ oscillations again around zero field after the strong field is withdrawn (similar to the low field measurement, $B$ is stepped down after +12 T polarization, see Fig. S15(A)). Then, similar $I_c(B)$ pattern is measured after applying -12 T, which also reproduces identical pattern ($B$ is stepped up after -12 T polarization, Fig. S15(B)). The robustness of the $I_c(B)$ pattern after applying +/-12 T are also reproduced in Device E ( Fig. S15(C), (D)). Therefore, the absence of neither hysteresis nor mirror symmetric pattern between up and down sweep with different strength of polarization field clearly shows that our data is drastically inconsistent with the ferromagnetic JJs, and thus we demonstrate beyond doubts that the observed φ-state is intrinsic to $K_2Cr_3As_3$ not related to ferromagnetism. We can also exclude the vortices as the source for the unconventional interference pattern by these three reasons, with an additional reason that we can reproduce the pattern after several cooling down processes with close-to-zero field.

We also note that neutron scattering measurements (*75*) identify the short-range spin polarization which flips on the atomic scale. Such spin structure with sub-nanometer scale will be averaged out in the micron-scale devices and cannot be the cause for our observed π-phase shift. A very weak internal field ~ 0.3 μT has also been found in $K_2Cr_3As_3$ by muon scattering measurement (*29*), which should also be short-ranged and the overall effect should be averaged out in our devices. Moreover, the SQUID oscillation has a periodic ~ 50 μT, more than two orders larger than the internal field of 0.3 μT. In Figs. 3(A) and 3(F), the weight of the 0-π components is adjusted by a magnetic field of ~2 mT, more than three-orders larger than the internal field. Any π-phase shift caused by 0.3 μT field is clearly not consistent with our observation where the magnetic modulation of the π phase shift happens in the scale of ~50 μT - 2 mT. Finally, even if we assume the very weak internal field ~ 0.3 μT as the coercive field of the possible ferromagnetic order, such scenario would result in a qualitatively different interference pattern (*76*). As illustrated in Fig. S16(A), so long as $B$ passes 0.3 μT, the spin polarization will be flipped and the $I_c(B)$ will switch to the branch of the opposite polarization. Since the coercive field is much smaller than the resolution of the field sweep, the hysteretic part in Fig. S16(A) is hardly visible (*76*). This will produce an overall mirror-symmetric $I_c(B)$, which is qualitatively different from our data in Fig. S16(B).

With the above evidences, we can confidently rule out the ferromagnetism, such as ferromagnetic impurities, the short-range magnetism, and the weak internal field in $K_2Cr_3As_3$ as the origins for our observed π-phase shift.

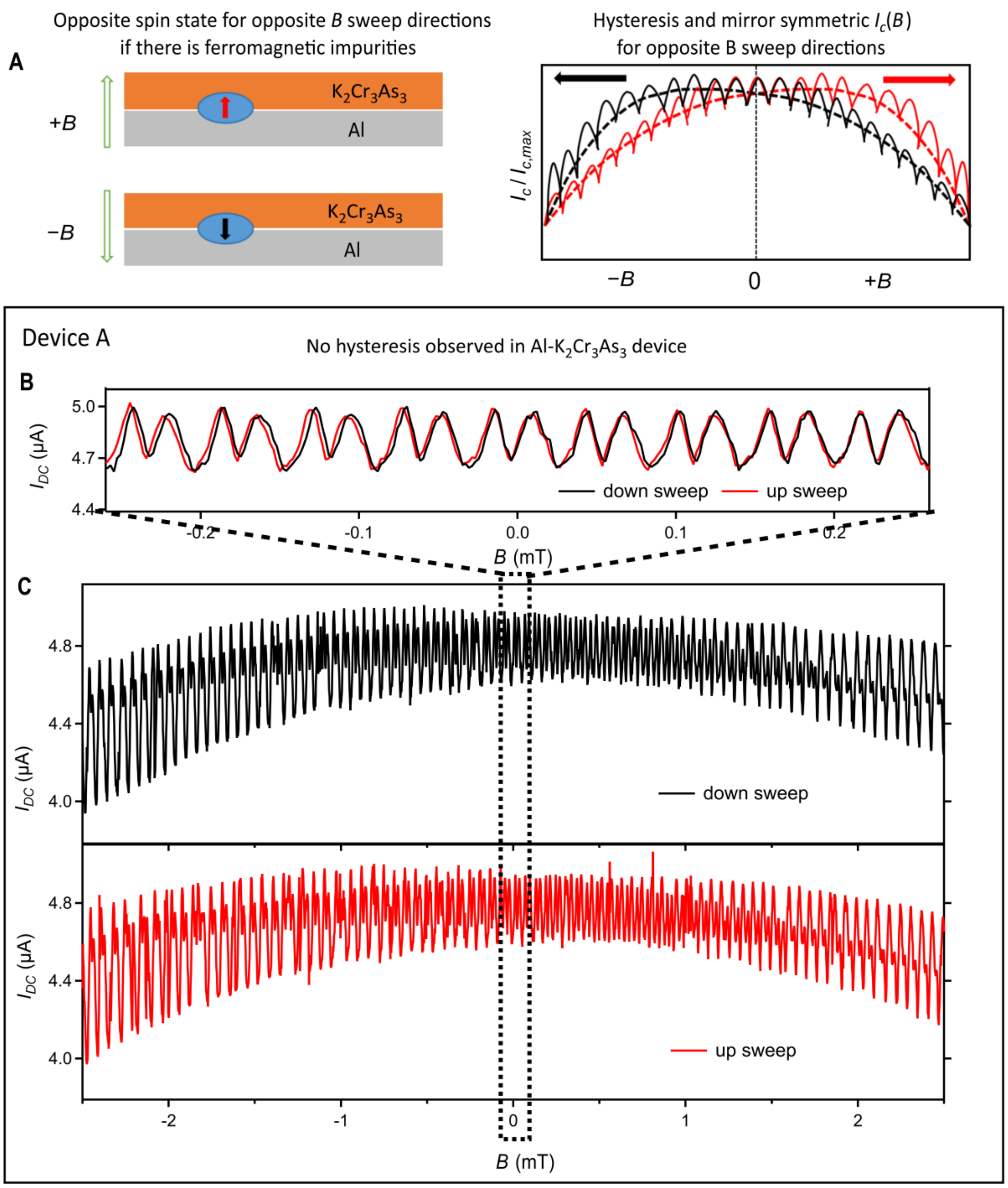

**Fig. S14. Ruling out ferromagnetism by showing no hysteresis or mirror-symmetric $I_c(B)$ in Al-$K_2Cr_3As_3$ SQUIDs. (A)** Cartoons (left) showing two equivalent states for a Josephson junction if there are ferromagnetic components with opposite spin polarizations by applying opposite magnetic field. Example $I_c(B)$ (right) if the SQUID contains ferromagnetic components in the opposite spin states, measured by stepping up (red) and down B (black). The Fraunhofer background (dashed lines) of the two $I_c(B)$ curves show hysteresis (*69*) and are mirror-symmetric with each other with respect to $B = 0$ (*70-72*). **(B)** Clear quasi-π-periodic $I_c(B)$ oscillations in Device A shows no hysteresis within ~ +/-0.3 mT. **(C)** Wider range sweep (+/-2.5 mT) shows neither hysteresis nor mirror-symmetric behaviors. All data are measured at 100 mK.

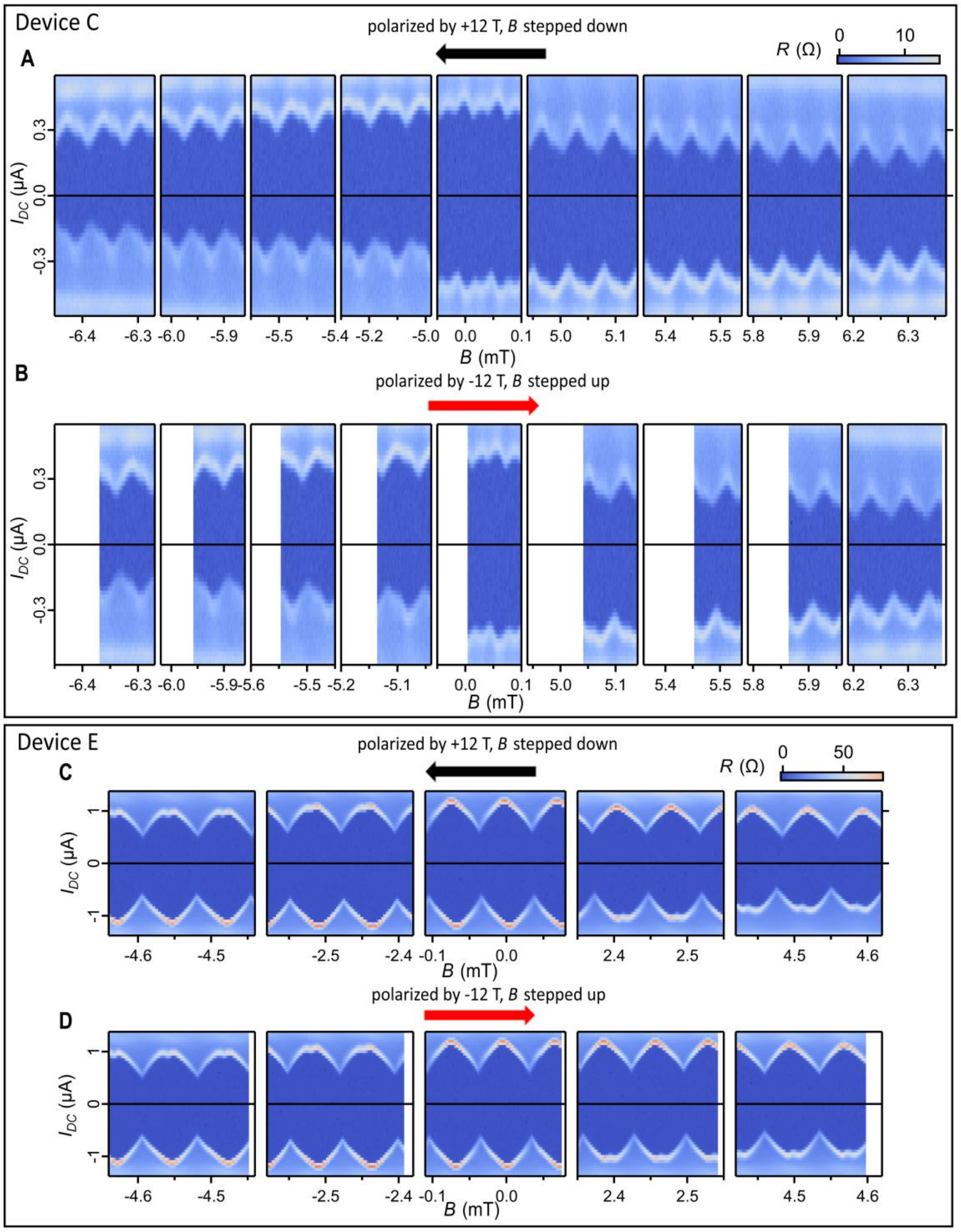

**Fig. S15. Ruling out ferromagnetism with coercive field up to +/-12 T in Al-$K_2Cr_3As_3$ SQUIDs. (A),(B)** $I_c(B)$ oscillations of Device C measured by stepping down (up) $B$, after being polarized with +12 T (-12 T) field. **(C),(D)** similar measurement of $I_c(B)$ oscillations in Device E. $T$ = 10 mK.

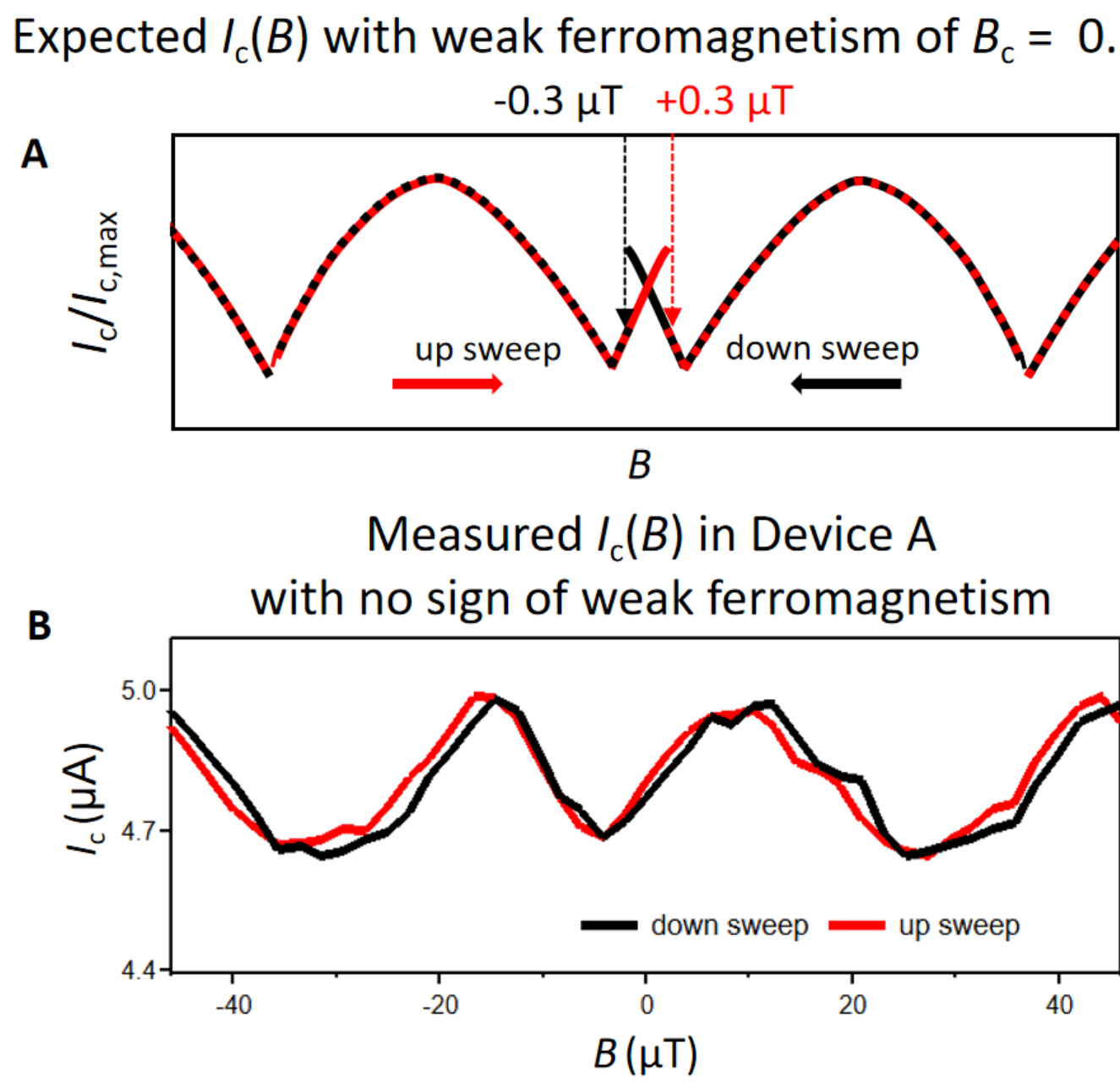

**Fig. S16. Disagreement between our data and a weak internal magnetism of 0.3 µT. (A)** Illustration of the expected $I_c(B)$ in the SQUID with a small coercive field of 0.3 µT. The up (red) and down sweeps (black) should also display symmetric pattern with $B$ = 0 (*76*). The section beyond the coercive field is marked as dashed lines. **(B)** Zoomed-in data $I_c(B)$ reproduced from Fig. S14(B) showing no symmetric pattern with $B$ = 0, inconsistent with the weak ferromagnetism.

## 12. Contribution of various segments to the SQUID transport

The SQUID used in the main text involves two kinds of superconductors, namely the $K_2Cr_3As_3$ stripes and the *s*-wave superconductor Al. As drawn in Fig. 1 and Fig. S17(A), there are three JJs: $JJ_{A,B}$ and the Al ring forming the SQUID, and $JJ_C$ in series with the SQUID. Both the SQUID ring and $JJ_C$ are further connected by Al to the external measurement circuit (contacts). In this section, we carefully characterize the transport features associated with each component. Fig. S17(B) is the $dV/dI(I_{DC}, B)$ in a wide range done at $T$ = 1.15 K. A horizontal blue band with the boundary marked by the brown arrow corresponds to the critical current of $K_2Cr_3As_3$. Indeed, $I_c \approx 15$ μA persists up to 10 T (Fig. S17(F)) and does not change up to $T$ = 1.4 K (Fig. S17(G)), in agreement with the previous work on the bulk crystal (*21,22*).

By zooming in the area at low bias (Fig. S17(C)), the regular oscillation is seen in the lower $I_c$ (blue arrow), which must come from the SQUID. The upper $I_c$ (black arrow) occurs at a finite-resistance state, indicating its origin as a component in series with the SQUID. The evolutions of the two $I_c$ in $dV/dI(I_{DC})$ with $T$ (Fig. S17(D)) trace closely the diminishing gap of Al up to $T$ = 1.2 K, which means they arise from the junctions $JJ_{A,B,C}$ with one superconducting electrode as Al. At a lower temperature $T$ = 0.05 K (Fig. S17(E)), the two $I_c$ are well separated. The lower $I_c$ clearly oscillates with flux while the upper $I_c$ does not oscillate at all, which unequivocally confirms that the lower $I_c$ belongs to the entire SQUID and the higher $I_c$ belongs to $JJ_C$ in series to the SQUID.

Another feature marked by the gray arrow in Fig. S17(B) has much higher $I_c$ up to 110 μA at zero field, see the inset) and a small $B_c$ (around 0.7 mT at zero bias). It also changes sensitively with $T < 1$ K, in agreement with the behavior of the bulk Al. We use the maxima of this $I_c$ to calibrate zero field because it is very sensitive to $B$. Meanwhile, from Fig. S17(C), such contribution is in series with the SQUID and $JJ_C$, which indicates that such contribution comes from the larger and further Al contacts connecting them. The fact that Al in the SQUID and the contacts have two separate $B_c$ and $I_c$ is possibly due to their different geometry. The different field profile near the $K_2Cr_3As_3$ stripes (see Fig. S3(B)) may also make the superconductivity of Al near $JJ_C$ more robust than that of the further away contacts.

In summary, the several features arising from $JJ_C$ and the Al contacts are in series with the SQUID, and thus do not affect the analysis of the $I_c$ oscillating with flux which is entirely determined by $JJ_{A,B}$ inside the SQUID.

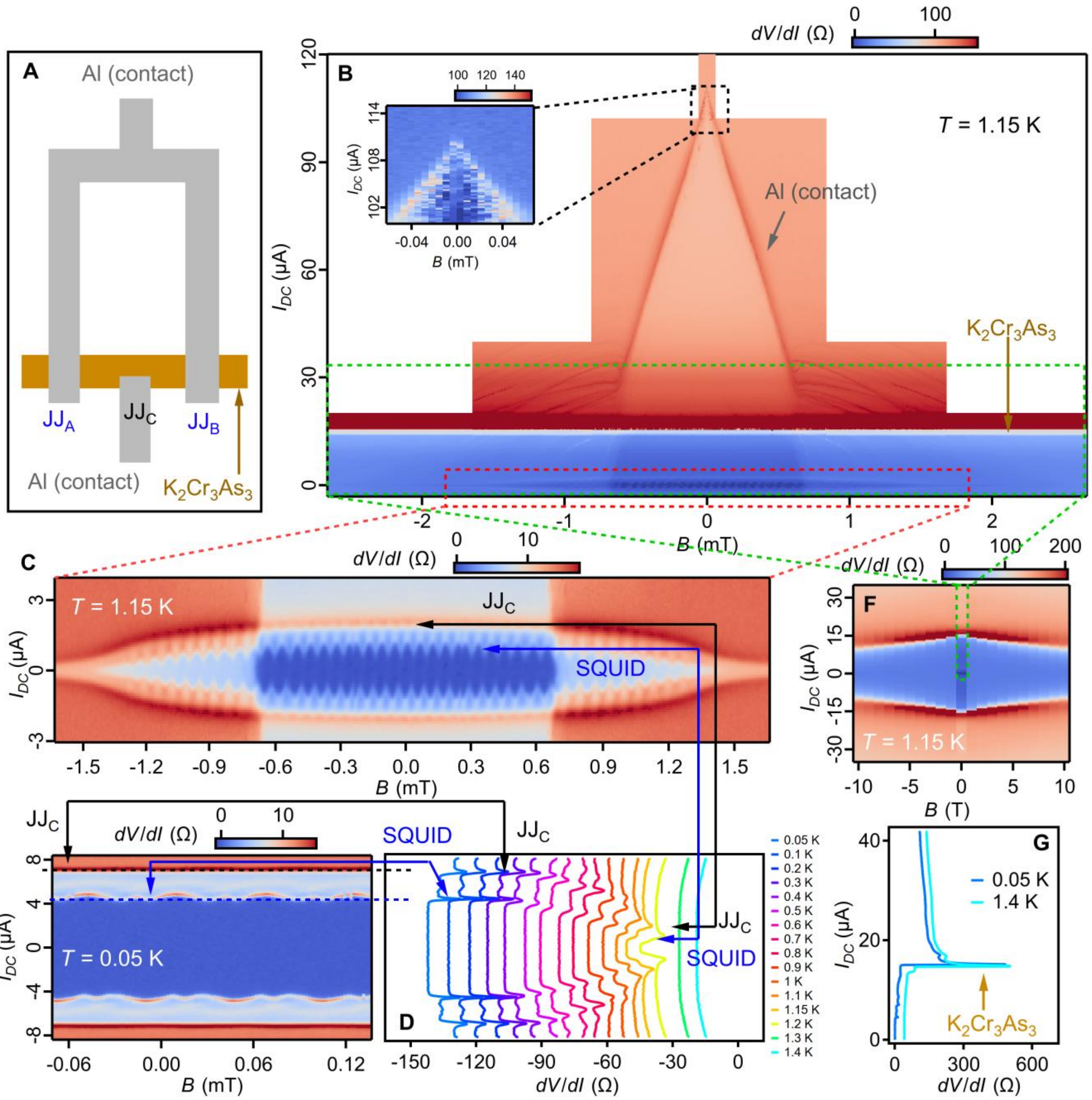

**Fig. S17. Contribution of various components to the SQUID transport. (A)** Schematic of Device A. **(B)** Wide range d$V$/d$I$($I_{DC}$, $B$) at $T$ = 1.15 K showing bulk $K_2Cr_3As_3$ (brown arrow) and Al critical currents (gray arrow). **(C)** Zoomed-in d$V$/d$I$($I_{DC}$, $B$) (red dashed rectangle) showing SQUID oscillation (blue arrow) and $JJ_C$ contribution in series (black arrow). **(D)** d$V$/d$I$($I_{DC}$) at $B$ = 0 and various $T$. The SQUID and $JJ_C$ contributions are in series and well separated at low $T$. **(E)** d$V$/d$I$($I_{DC}$, $B$) at $T$ = 0.05 K with clearly separated $I_c$. The one from $JJ_C$ shows no SQUID oscillations. **(F)** High field d$V$/d$I$($I_{DC}$, $B$) showing $K_2Cr_3As_3$ supercurrent up to 10 T. The green dashed rectangle contains (B) data. **(G)** Wide range d$V$/d$I$($I_{DC}$) at $B$ = 0 showing $I_c$ of $K_2Cr_3As_3$ insensitive to $T$ up to 1.4 K.

## 13. Half-integer Shapiro steps in φ-junction

RSJ model justified by overdamped JJ

Figure S18 shows the measurements of Device A where the $I_{DC}$ is swept from superconducting to normal states (0-p, Fig. S18(A)) and from normal to superconducting states (p-0, Fig. S18(B)). The transition current is the switch current $I_c$ and retrap current $I_r$ respectively. Fig. S18(C) shows a typical linecut, demonstrating $I_c = I_r$ and thus the overdamped scenario (*68*) in our SQUIDs. This greatly simplifies the analysis of the phase dynamics under microwave irradiation, and it is justified to use the RSJ model (*68*).

Considering the SQUID as a whole system with the flux-tunable CPR $I_{tot}(\varphi, \Phi)$, where the flux Φ through the SQUID is fixed when calculating the Shapiro steps. Therefore, for the phase drop $\varphi$ across the whole SQUID, we have (*39*):

$$\frac{d\varphi}{d\tau} + i_{\text{tot}}(\varphi, \Phi) = i_{\text{DC}} + i_{\text{AC}} \cos(2\pi\nu\tau) \quad \text{(S3)}$$

where $i_{tot} = I_{tot}/I_C$, $i_{DC} = I_{DC}/I_C$, and $i_{AC} = I_{AC}/I_C$ are SQUID supercurrent, the DC bias current, and the AC current induced by the microwave respectively, all normalized by the critical current of the SQUID. The Josephson frequency of the system is $f_J = 2eI_cR/h$ where $R$ is the normal resistance. In equation (S3), $\nu = f/f_J$ and $\tau = f_J t$ are the normalized irradiation frequency and time.

Modeling $i_{tot}(\varphi) = I_1/I_c \sin(\varphi) - I_2/I_c \sin(2\varphi)$, the Shapiro map of the averaged differential resistance $dV/dI(i_{AC}, i_{DC})$ is determined by $g = |I_1/I_2|$ and $\nu$. In Fig. 4(A) (reproduced in Figs. S19(A)), we show that for $g = 1.2 > 1/2$, the Shapiro map exhibits the 1/2 Shapiro step (red arrows) between sequential integer Shapiro steps. In Figs. S19(B) with the same $\nu$ but with $g = 1/2$, only integer steps (white arrow) are seen.

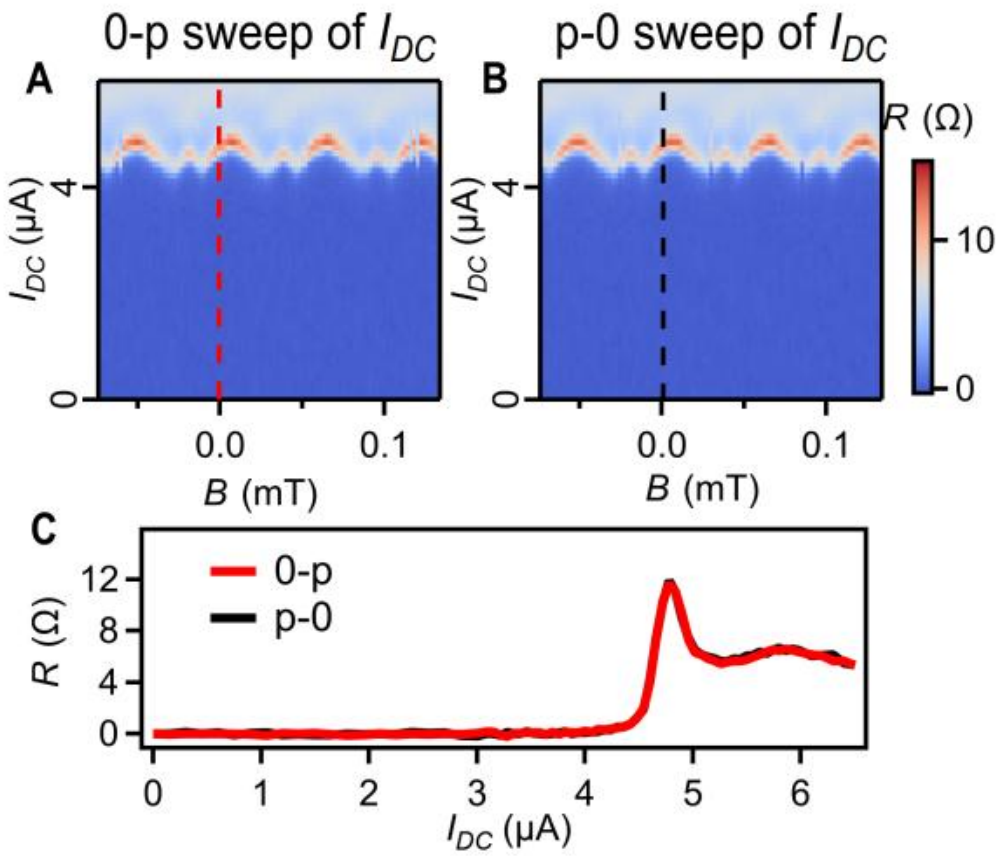

**Fig. S18. Overdamped JJs demonstrated by equal switch and retrap current. (A), (B)** d$V$/d$I$($I_{DC}$, $B$) for 0-p and p-0 sweeps, measuring the switch and retrap critical currents, respectively. **(C)** A typical d$V$/d$I$($I_{DC}$) taken from the same $B$ in (A), (B). No hysteresis is observed.

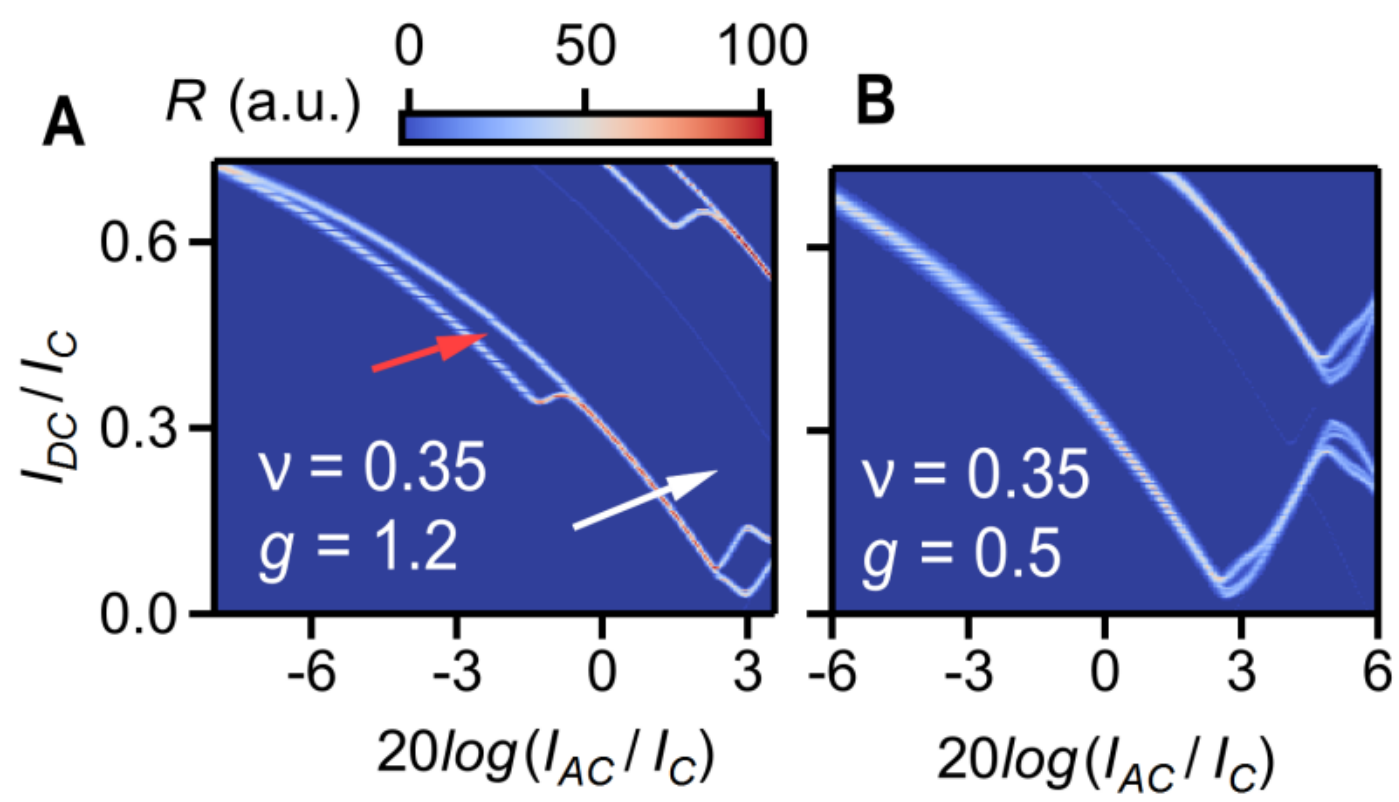

**Fig. S19. $\varphi$-JJ state revealed by the half-integer steps. (A), (B)** Calculated $dV/dI(I_{DC}/I_C, 20\log(I_{AC}/I_C))$ by equation (S3). $\nu = f/f_J = 0.35$ same as the main text Fig. 4. $g = |I_2/I_1| = 1.2$ (A) and 0.5 (B), respectively. The half-integer steps (red arrows) between the integer steps (white arrows) before the first node are only visible for $g > 1/2$.

Shapiro step measuremens in Device B

Fig. S20 shows the microwave response for $\nu = f/f_J$ = 0.35 of Device B shown in Fig. S10. Figs. S20(A)-(C) are the same as Figs. 4(H)-(J). Interestingly, in the field range where $I_{c-}$ shows the quasi-π-periodic oscillations, $I_{c+}$ only shows the 2π-periodic oscillations. There, we reproduce the clear and continuous 1/2 step (Figs. S20(B), (C) (red arrows)) also only for the negative $I_{DC}$. The Shapiro maps (Figs. S20(F), (G)) are taken at $B$ indicated by the vertical dashed lines in Figs. S20(A)-(C). In all of the Shapiro maps, a clear 1/2 step (red arrows, highlighted by the $V$-$I$ curves in Figs. S20(D), (E)) similar to Fig. S20(A) shows up only for the negative $I_{DC}$ with quasi-π-periodic feature and disappears for the positive $I_{DC}$ with 2π-periodic feature. These observations further confirm that a robust and dominant second harmonics is the source for both the quasi-π-periodic feature and the 1/2 Shapiro step for the negative $I_{DC}$.

We note that although the half-integer steps may appear in quasi-ballistic JJs for larger frequencies and power (*39*), for the low frequency (much lower than $f_J$) and power (for example, before the first node) used in our experiment, the emergence of 1/2 step can be used as an indicator for the $g$ > 1/2 in our system (*36-38*,*43*). In the main text, we also exclude other alternative mechanisms such as interference effect (*54*,*55*) and non-equilibrium (*53*). In particular, the comparison between positive and negative bias in Device B clearly shows that This further eliminates the interpretation of the quasi-π-periodic $I_c(B)$ as the beating between two 2π-periodic oscillating channels (*50*), since the beating effect alone does not induce large second harmonics with $g$ > 1/2. Therefore, the measured quasi-π-periodic $I_c(B)$, as well as the robust half-integer Shapiro steps, can only be caused by the co-existence of 0- and π-phases supercurrent.

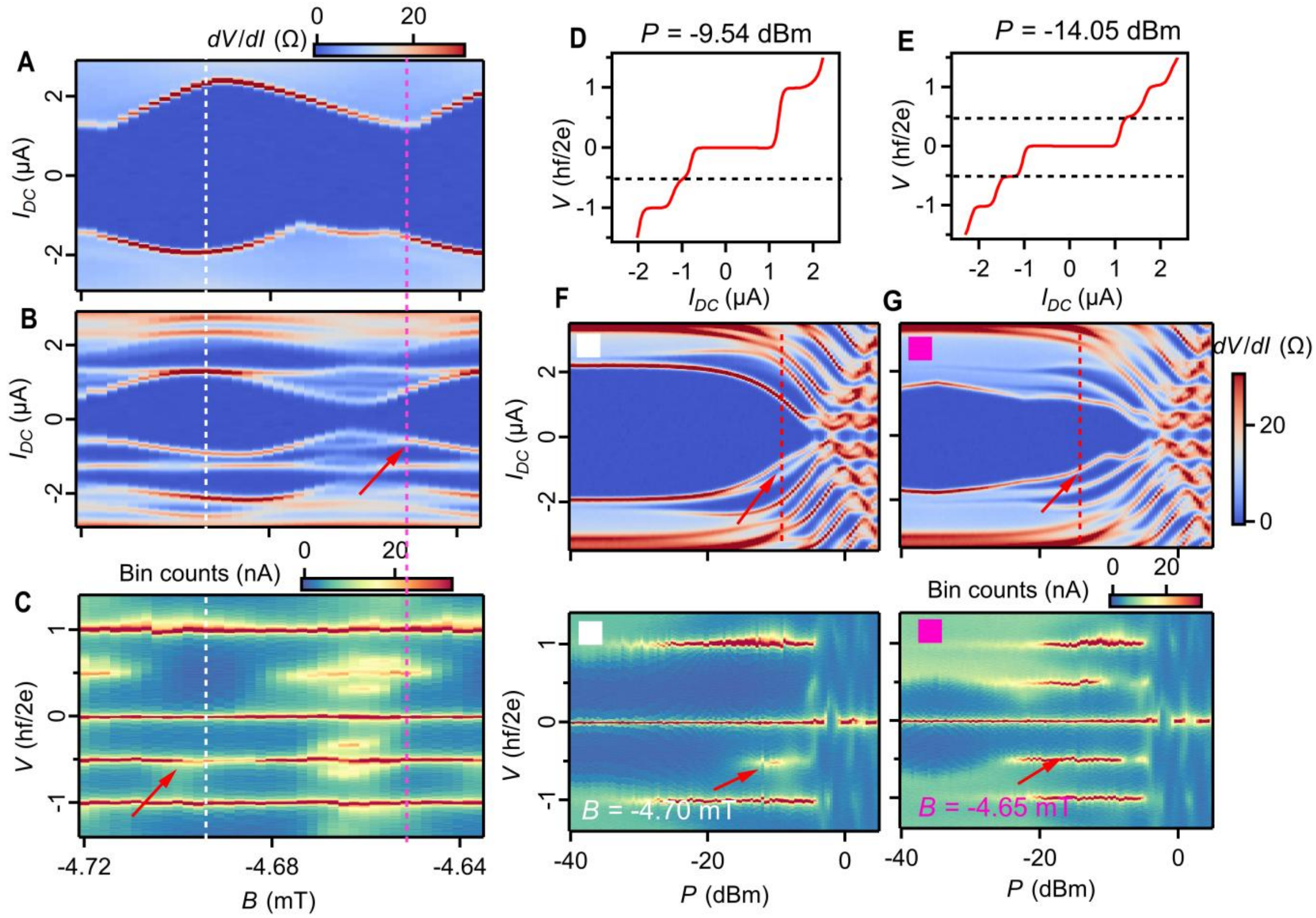

**Fig. S20. Device B reproducing continuous 1/2 Shapiro step in negative bias. (A)** d$V$/d$I$($I_{DC}$, $B$) without microwave at $T$ = 96 mK, showing quasi-π-periodic oscillations for negative bias and 2π-periodic oscillations for positive bias. **(B), (C)** d$V$/d$I$($I_{DC}$, $B$) and the converted histograms with the irradiation of $f$ = 3.38 GHz (or $\nu$ = 0.35, $I_c$ = 2.0 μA， $R_N$ = 9.9 Ω) and $P$ = -10 dBm. The 1/2 step (red arrow) is continuous over the whole period only for the negative bias. **(D)** $V$-$I$ curve at $P$ = -9.56 dBm and $B$ = -4.696 mT (integration of the red dashed line $R$-$I$ in (F)), showing 1/2 step (dashed line). **(E)** Similar $V$-$I$ curve at $P$ = -14.05 dBm and $B$ = -4.646 mT marked by the red dashed line in (G). **(F)** d$V$/d$I$($I_{DC}$, $P$) (upper) and the converted histogram (lower) at $B$ = -4.696 mT (the white vertical dashed lines in (A)-(C)), confirming the robust 1/2 step (red arrows). **(G)** Similar to (F) but with $B$ = -4.646 mT (the magenta vertical dashed lines in (A)-(C)), all showing clear 1/2 step.